\documentclass[9pt,prx,letterpaper,nobalancelastpage,twocolumn,superscriptaddress,nofootinbib]{revtex4-2}

\usepackage{newtxtext, newtxmath} 
\usepackage{geometry}
\usepackage[colorlinks = true,]{hyperref}
\usepackage{float}
\usepackage{graphicx}
\usepackage{dcolumn}
\usepackage{bm}
\usepackage[dvipsnames]{xcolor}
\usepackage{braket}
\usepackage{soul}
\usepackage{comment}
\usepackage{amsmath}
\usepackage{cases}
\usepackage{mathtools}
\usepackage{xcolor}
\usepackage[normalem]{ulem}

\newcommand{\crea}[1]{{\hat{#1}}^\dagger}
\newcommand{\ann}[1]{\hat{#1}}
\newcommand{\kerr}[1]{\hat{#1}^\dagger \hat{#1}^\dagger \hat{#1} \hat{#1}}
\newcommand{\round}[1]{ \left( #1 \right) }
\newcommand{\squared}[1]{ \left[ #1 \right] }
\newcommand{\curly}[1]{ \left\{ #1 \right\} }
\newcommand{\trace}[1]{\text{Tr}}

\DeclareMathOperator{\Tr}{Tr}
\DeclareMathOperator{\PV}{PV}

\begin{document}

\title{Nonlocal Quantum Phase Transitions}

\author{Alessandro Coppo}
\email{alessandro.coppo@cnr.it}
\affiliation{Institute for Complex Systems, National Research Council (ISC-CNR), Via dei Taurini 19, 00185 Rome, Italy}
\author{Aanal Jayesh Shah $ ^\ddagger$}
\affiliation{Department of Physics and Astronomy, Purdue University, West Lafayette, IN 47906, USA}
\author{Hadiseh Alaeian}
\affiliation{Elmore Family School of Electrical and Computer Engineering, Purdue University, West Lafayette, IN 47906, USA}
\affiliation{Department of Physics and Astronomy, Purdue University, West Lafayette, IN 47906, USA}
\author{Valentina Brosco}
\affiliation{Institute for Complex Systems, National Research Council (ISC-CNR), Via dei Taurini 19, 00185 Rome, Italy}
\affiliation{Physics Department, Sapienza University, P.le A. Moro 2, 00185 Rome, Italy}
\author{Roberto Di Candia}
\affiliation{Department of Information and Communications Engineering, Aalto University, Espoo 02150, Finland}
\affiliation{Dipartimento di Fisica, Università degli Studi di Pavia, Via Agostino Bassi 6, I-27100 Pavia, Italy}
\author{Simone Felicetti}
\email{simone.felicetti@cnr.it}
\affiliation{Institute for Complex Systems, National Research Council (ISC-CNR), Via dei Taurini 19, 00185 Rome, Italy}
\affiliation{Physics Department, Sapienza University, P.le A. Moro 2, 00185 Rome, Italy}

\maketitle
\footnotetext[0]{$\!\!\!\!^\ddagger$ Co-first author: A.C. contributed to the analytical theory and the general setting; A.J.S. performed the numerical simulations.}
\noindent\textbf{
Phase transitions are paradigmatic examples of emergent phenomena, in which symmetries present at the microscopic level can be spontaneously broken in the thermodynamic limit. Two primary physical mechanisms can drive this symmetry breaking: \emph{thermal} fluctuations in classical phase transitions and \emph{quantum} fluctuations in quantum critical phenomena.
Here, we introduce \emph{nonlocal} quantum fluctuations as a new fundamental mechanism to drive phase transitions. We show that entanglement shared between environmental modes can induce a correlated symmetry breaking in remote systems, independent of their spatial separation. Using the framework of driven–dissipative phase transitions, we theoretically investigate a system composed of two nonlinear quantum resonators placed at arbitrarily large spatial separations, each coupled to independent local Markovian baths. We consider the regime in which remote environmental modes are prepared in broadband entangled states. We show that near the critical point, where the susceptibility to weak perturbations diverges, quantum correlations in the environments 
govern the system's critical behavior. While these correlations manifest locally only as effective thermal fluctuations, at the global level they give rise to an emergent nonlocal phase transition, marked by the spontaneous symmetry breaking of a collective mode shared by the two remote systems.}

The study of phase transitions is relevant 
across different research areas such as condensed matter, statistical mechanics, complex systems, cosmology, and high-energy physics. In the context of quantum information, critical phenomena are also considered a conceptual framework and a practical resource for quantum computing~\cite{Ad_q_comp} and metrology~\cite{montenegro_review_2025,mihailescu2025tutorial}. Phase transitions are usually studied in systems at thermal equilibrium with their environment. While classical phase transitions are driven by thermal fluctuations~\cite{Huang1987}, quantum phase transitions 
 persist at zero temperature~\cite{Sachdev2011}.
The concept of thermal and quantum phase transitions can be extended to systems driven out of thermal equilibrium by an external source. Driven-dissipative systems can indeed exhibit critical phenomena~\cite{Maghrebi16, Minganti2018}, where the non-equilibrium steady-state manifold undergoes a nonanalytic change in response to an infinitesimal variation of a control parameter. This behavior has been experimentally observed in a wide variety of atomic~\cite{baumann_dicke_2010, nagy_dicke-model_2010, klinder_dynamical_2015, Ferri2021, cai2021observation, helson_density-wave_2023, song_dissipation-induced_2025} and solid-state systems~\cite{RodriguezPRL17, FitzpatrickPRX17, FinkPRX17, FinkNatPhys18, Brookes2021, Zejian2022, sett2022emergent, Chen2023, beaulieu_observation_2025}, and has also been applied in proof-of-concept sensing experiments~\cite{ding2022enhanced, petrovnin2023microwave, beaulieu_criticality-enhanced_2025}.  

Critical phenomena typically have been studied in the thermodynamic limit of many-body systems, where the number of constituents tends to infinity. 
However, non-analyticities can also emerge in finite-component systems, provided that they can explore an infinite-dimensional Hilbert space~\cite{hwang_quantum_2015, felicetti2020universal}. Here, the conventional thermodynamic limit is replaced by an asymptotic rescaling of system parameters, which triggers a macroscopic growth of excitations. Nonlinear quantum resonators provide a paradigmatic and physically relevant example, where first- and second-order dissipative phase transitions have recently been observed with trapped ions~\cite{cai2021observation} and superconducting circuits~\cite{Chen2023, beaulieu_observation_2025}. Such platforms enable the study of critical phenomena in a highly controllable way, and they are of high interest for quantum-computing applications~\cite{grimm_20}. Furthermore, both atomic and solid-state systems can interact with entangled signals supported by optical or telecom photons~\cite{van_leent_entangling_2022,krutyanskiy_entanglement_2023, knaut_entanglement_2024, ruskuc_multiplexed_2025} or propagating microwaves~\cite{magnard_microwave_2020,fedorov_experimental_2021, casariego_propagating_2023, storz_loophole-free_2023, yam_cryogenic_2025, juanes_entangling_2025}. 

\begin{figure*}[ht]
    \centering
    \includegraphics[width=1\textwidth]{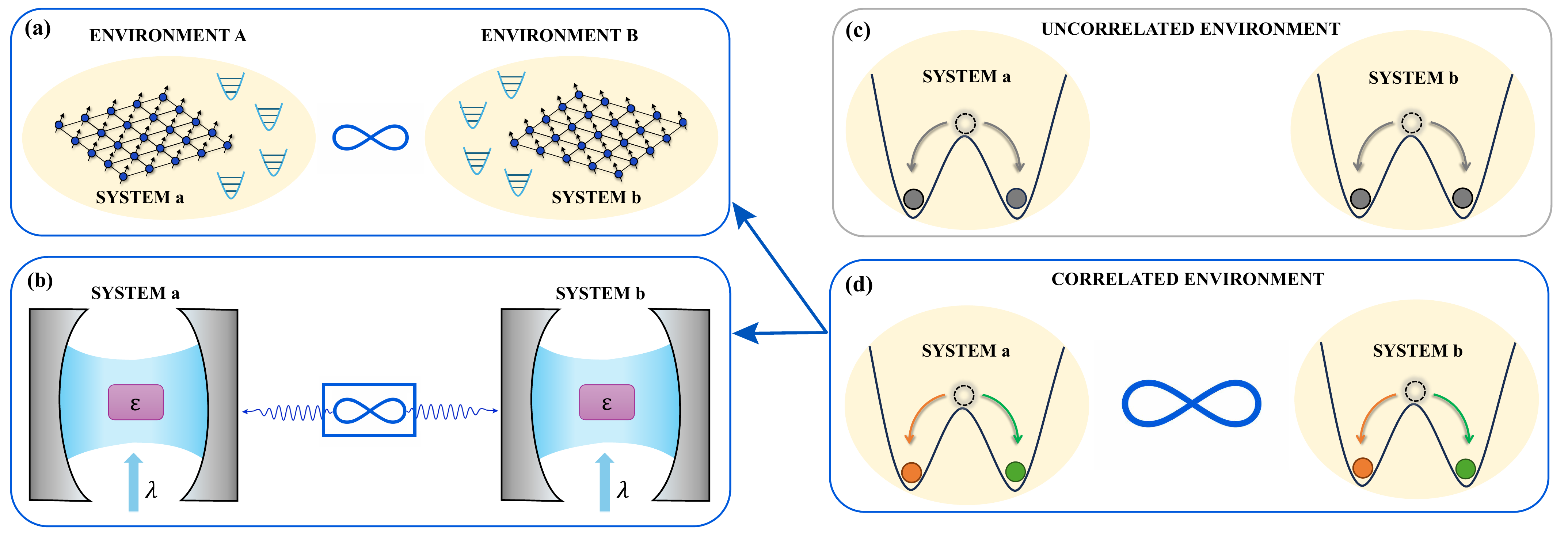}
    \caption{\textbf{Sketch of the general framework.} (a) Two arbitrarily remote critical quantum systems, represented by generic spin models, each coupled to the respective environment. The two environments are prepared in an entangled state, but neither the systems nor the environments interact with each other. (b) The particular setup considered in this work: two distant critical Kerr resonators whose environments are prepared in a broadband two-mode squeezed vacuum state. (c)-(d) Independent and collective spontaneous symmetry breaking in the double-well picture. In the independent (uncorrelated) case (c), each particle localizes in either the left or the right well independently. In the collective case (d), environmental quantum correlations induce joint localization, with both particles collapsing in the same well.}
    \label{fig:cartoon}
\end{figure*}

In this work, we introduce a novel conceptual framework in which phase transitions are induced by nonlocal quantum fluctuations. As a minimal model, we consider a driven-dissipative quantum nonlinear resonator embedded in a Markovian bath. Besides its direct experimental relevance~\cite{Chen2023, beaulieu_observation_2025}, this model is known to effectively reproduce the critical phenomena of a broad class of fully-connected models~\cite{Garbe22}, such as the infinite-range Ising, Rabi, and Dicke models. We theoretically study the physical setting in Fig.~\ref{fig:cartoon}, where two identical copies of such critical systems are placed at an arbitrarily large separation, each interacting with its own bath. The baths are not directly coupled, but their modes are assumed to be in an entangled state. Using the Caldeira-Leggett model of bosonic environments, we derive the Lindblad master equation~\cite{BreuerBookOpen} for a broadband two-mode squeezed state of the baths~\cite{SerafiniBOOK2017quantum}. We then characterize the global system's critical behavior using complementary analytical and numerical methods, each revealing distinct features of the nonlocal phase transition.

As a first step, we develop a Gaussian theory, which includes second-order (Bogoliubov) quantum fluctuations on top of mean-field solutions. This approach describes the driven-dissipative phase diagram of a single critical resonator, far  from the critical region~\cite{Carusotto_RMP_2013_quantum_fluids_light, Minganti2018, di2023critical}. This Gaussian approximation can also capture the two-mode squeezing of Bogoliubov excitations, due to the nonlocal bath fluctuations. However, the considered nonlocal bath correlations do not change the mean-field solutions nor the structure of the symmetry-broken phases. This suggests that nonlocal correlations do not modify the system’s critical behavior---a seemingly negative result. 

Interestingly, while a Gaussian model does not predict any modification to the symmetry breaking, exact full-quantum numerical simulations show nontrivial properties of the global-system steady state in the vicinity of the critical point. A \emph{local} observer could only witness a standard driven-dissipative phase transition in the presence of a thermal bath. Yet, an observer with access to the \emph{global} system could resolve the emergence of delocalized collective modes, for which the phase transition is enhanced or suppressed. Numerical evidence indicates that, owing to the nonlocal character of the transition, the two critical resonators undergo correlated symmetry breaking despite being uncoupled and separated by arbitrarily large distances.

To understand the emergence of this nonlocal symmetry breaking, we develop a non-Gaussian theory valid in the critical region. We use phase-space methods and build on the slaving principle~\cite{hutt_synergetics_2020,dykman2012fluctuating}, which leverages the separation of time scales induced by critical slowing down. 
Our analytical approach unveils the emergence of \emph{nonlocal} critical modes induced by nonlocal quantum fluctuations, and it provides a clear interpretation of their physical origin. Overall, these analytical and numerical results open a new research direction centered on critical phenomena induced by nonlocal quantum fluctuations, including their experimental realization and application in quantum information science.

\section{General framework}

One of the main contributions of this work is to introduce the general framework sketched in Fig.~\ref{fig:cartoon}. To introduce it, let us consider first a single, closed critical quantum system.
For simplicity, we focus on the case in which criticality arises from $\mathbb{Z}_2$ spontaneous symmetry breaking, where the Hamiltonian commutes with the parity operator $\hat{\Pi}$. As a control parameter $\lambda$ is varied across a critical value $\lambda_c$, the system undergoes a phase transition. For $\lambda < \lambda_c$, the ground state $|0\rangle$ is unique and symmetric, satisfying $\hat{\Pi}|0\rangle = |0\rangle$. For $\lambda > \lambda_c$, the low-energy sector develops two quasi-degenerate parity eigenstates, resulting from the superposition of the two symmetry-broken states $|R\rangle$ and $|L\rangle$, related by $\hat{\Pi}|R\rangle = |L\rangle$. In the thermodynamic limit the degeneracy becomes exact, and the system spontaneously settles into either $|R\rangle$ or $|L\rangle$. As a result, the ground state no longer preserves the Hamiltonian symmetry. These two phases can be characterized by an order parameter, defined as the expectation value of an observable $\hat{O}$ anticommuting with $\hat{\Pi}$. The order parameter vanishes in the normal phase $\lambda<\lambda_c$, whereas, in the symmetry-broken phase $\lambda>\lambda_c$, it acquires a finite value, with opposite sign in the states $\ket{L}$ and $\ket{R}$. This mechanism is often illustrated through the picture of a particle in an effective double-well potential.

Similar critical phenomena arise in non-equilibrium open quantum systems, where phase transitions occur in the steady state $\rho_a$ reached asymptotically under the combined effect of external driving and dissipation~\cite{Minganti2018}. In the normal phase, the steady state $\rho_0$ preserves the symmetry, yielding a vanishing order parameter, $\Tr[\rho_0 \hat{O}]=0$. Beyond the critical point, in the thermodynamic limit, the steady state becomes a classical mixture of two symmetry-broken states $\rho^L$ and $\rho^R$. For finite-size systems, these states are metastable, and continuous monitoring reveals quantum jumps between them. In the thermodynamic limit, the switching rate vanishes asymptotically~\cite{Minganti2018}, and the system remains indefinitely in either $\rho^L$ or $\rho^R$. The steady state is thus described by the balanced mixture $\rho = (\rho^L + \rho^R)/2$, which must be interpreted as a statistical average over many independent experimental realizations. Notably, the expectation value $\Tr[\rho \hat{O}]$
vanishes. Nevertheless, the density operator $\rho$ corresponds to a symmetry-broken state since the result of any single-shot measurement of $\hat{O}$ yields a \emph{macroscopic} finite value of the order parameter.\\
Consider now two identical copies, $a$ and $b$, of a critical quantum system, each one coupled to its own local quantum environment. Neither the systems nor their environments interact with each other. When the environments are uncorrelated, in the symmetry-broken phase the global system collapses with equal probability into one of the four symmetry-broken states $\rho^L_{a}\otimes\rho^L_{b}\,,\, \rho^L_{a}\otimes\rho^R_{b}\,,\, \rho^R_{a}\otimes\rho^L_{b}\,,\, \rho^R_{a}\otimes\rho^R_{b}$. The global steady state reads $\rho = (\rho^L_{a} + \rho^R_{a})/2 \otimes (\rho^L_{b} + \rho^R_{b})/2$. In the double-well picture, each particle independently settles into either the left or the right minimum of its respective potential, as depicted in Fig.~\ref{fig:cartoon}(c).

We can finally introduce the central setup of this work.
While the two critical systems and their respective environments remain spatially remote and strictly uncoupled, the environments are now prepared in an entangled state. This configuration raises fundamental questions: can quantum correlations between the environmental states actively modify the structure of the symmetry-broken phase and, more broadly, the very nature of the phase transition itself? Can they give rise to a non-locally ordered phase where the two systems become either correlated, $\rho = (\rho^{L}_{a} \otimes \rho^{L}_{b} + \rho^{R}_{a} \otimes \rho^{R}_{b})/2$, or anti-correlated, $\rho = (\rho^{L}_{a} \otimes \rho^{R}_{b} + \rho^{R}_{a} \otimes \rho^{L}_{b})$? In the double-well picture, this non-local order would mean that the remote systems no longer break the symmetry independently. Rather, the environmental quantum correlation perfectly locks their states, driving the particles to collapse jointly into either the same or strictly opposite wells, even in the complete absence of a direct coupling, as illustrated in Fig.~\ref{fig:cartoon}(d). 

\section{Model derivation}
\label{sec_model_derivation}
We focus now on the system composed of driven Kerr resonators of Fig.~\ref{fig:cartoon}(b), which are minimal critical systems whose dissipative phase transitions have recently been experimentally characterized~\cite{Chen2023,beaulieu_observation_2025}. 
We consider a two-photon drive and write the Hamiltonian in the frame rotating at the drive frequency (see SI),
\begin{equation}
\label{Kerr}
\hat{H}_a = \omega \crea{a}\ann{a} + \frac{\lambda}{2} \round{\crea{a}\crea{a} + \ann{a}\ann{a}} + \epsilon \crea{a}\crea{a}\ann{a}\ann{a},
\end{equation}
where $\omega$ is the drive-to-cavity detuning, $\lambda$ is the two-photon drive intensity, and $\epsilon$ is the Kerr nonlinearity. This model becomes critical in the limit $\epsilon\to 0$, which plays the role of an effective thermodynamic limit. Equation \eqref{Kerr} reproduces indeed the critical behavior of the fully-connected Ising model~\cite{Garbe22,Dusuel05} up to corrections of order $\mathcal{O}(1/N^2)$, under a re-parametrization in which $\epsilon \sim 1/N$, where $N$ is the number of spins (see Fig.~1(a) and SI). We assume the resonator is coupled to a Caldeira-Leggett bath ~\cite{BreuerBookOpen}, composed of a continuum of bosonic modes with free Hamiltonian 
$\hat{H}_A = \int d\omega\  \omega \crea{A}_\omega \ann{A}_\omega$, where $\ann{A}_\omega$ are annihilation operators of the spectral modes of the environment $A$. The system-bath coupling can be written as $\hat{H}_{Aa} \! = \!\! \int \! d\omega \frac{g(\omega)}{\sqrt{2 \pi}} \round{\crea{a} + \ann{a}}\round{\crea{A}_\omega + \ann{A}_\omega}$, where $g(\omega)$ is a coupling function proportional to the bath's density of states. The total Hamiltonian for this system-environment model is $H_A^{SE} = H_a + H_A + H_{Aa}$. Following the standard procedure for a zero-temperature Markovian bath, we get the  master equation for the reduced density matrix of system $a$,
\begin{equation}
\label{Lindblad_a}
\partial_t{\rho}_a = \mathcal{L}_a(\rho_a) =- i\left[\hat{H}_a,\rho_a(t)\right] + \gamma\, \mathcal{D}[\ann{a}](\rho_a),
\end{equation}
where $\mathcal{D}[\hat{O}](\cdot) = 2 \hat{O} \cdot \hat{O}^\dagger - \{\hat{O}^\dagger\hat{O},\cdot\}$ describes the cavity photon decay process at rate $\gamma$. It is well known that this model exhibits a continuous dissipative phase transition in the limit $\epsilon\to 0$, as the drive strength $\lambda$ is increased~\cite{Minganti2018, beaulieu_observation_2025}.
Below the critical driving strength $\lambda < \lambda_c$, the steady state is a squeezed vacuum. Above the threshold $\lambda > \lambda_c$, it is given by the statistical mixture $\rho_a = \frac{1}{2}(\rho_a^{\alpha} + \rho_a^{-\alpha})$, where $\rho_a^{\alpha} = \ket{\alpha}\bra{\alpha}_a$ are coherent states breaking the $\mathbb{Z}_2$-symmetry given by the parity operator $\hat{\Pi}=e^{i\pi \crea{a}\ann{a}}$. The order parameter can be identified with the real part of the displacement $\alpha$, which diverges as $|\alpha|^2 \sim 1/\epsilon$.

This sets the ground to introduce the complete model of Fig.\ref{fig:cartoon}(b), composed of two local copies of this setup. The total Hamiltonian is expressed as $H_\textrm{tot} = H_A^{SE} + H_B^{SE}$, with $\hat{b}$ and $\hat{B}_\omega$ denoting the annihilation operators for system $b$ and its environment, respectively. 
Non-locally-correlated environments are modeled by a broadband two-mode squeezed state satisfying $\langle\crea{A}_\omega \ann{A}_{\omega^\prime}\rangle = \langle\crea{B}_\omega \ann{B}_{\omega^\prime}\rangle = \delta(\omega-\omega^\prime)\,n_r$, 
$\langle\crea{A}_\omega \crea{B}_{\omega^\prime}\rangle = \delta(\omega-\omega^\prime)\,m_r$. The parameters $n_r = \sinh^2{r}$ and $m_r= -e^{i\phi} \sinh{r}\cosh{r}$ are defined in terms of the two-mode squeezing strength $r$ and phase $\phi$.
We assumed that the spectrum of squeezing is flat over the relevant frequency band. Under standard Markovian approximation~\cite{BreuerBookOpen,cohen2024Book}, we  microscopically derived a Lindbladian master equation for the global system (see SI),
\begin{equation}
\label{Lindblad_ab}
\partial_t \rho = \mathcal{L}_a(\rho) + \mathcal{L}_b(\rho) + \mathcal{L}_{\rm sq}(\rho),
\end{equation}
where $\rho$ is the joint density matrix of the two resonators.
The bath-mode squeezing contributes to the master equation with local ($n_r$) and nonlocal ($m_r$) contributions,
\begin{align}
\label{Lindblad_sq}
\mathcal{L}_{\rm sq}(\rho) &= \gamma\,n_r \Big(\mathcal{D}[\crea{a}](\rho) + \mathcal{D}[\crea{b}](\rho)+\mathcal{D}[\ann{a}](\rho) + \mathcal{D}[\ann{b}](\rho)\Big) \nonumber \\
&+ 2\gamma\,m_r \left(\ann{a}^\dagger\rho \hat{b}^\dagger +\hat{b}^\dagger\rho \hat{a}^\dagger-\{\hat{a}^\dagger \hat{b}^\dagger,\rho\} + \text{H.c.} \right) ~.
\end{align}
The first line contains only local contributions and is formally equivalent to the effect of a finite-temperature bath. Indeed, locally, a two-mode squeezed state is indistinguishable from a thermal field with $n_r$ excitations. The second line, by contrast, encodes genuinely nonlocal quantum fluctuations: it vanishes upon tracing over either subsystem and therefore does not affect local observables. These nonlocal fluctuations $m_r$ are accessible only through global observables requiring joint measurements on both subsystems, and they are at the origin of the novel phenomenology explored in this work.

\section{Gaussian theory}
In this section, we derive analytical solutions for the system steady state under Gaussian approximation, valid in the effective thermodynamic limit $\epsilon\to0$ and far from the critical region around $\lambda_c$. We rewrite the field modes as $\ann{a}=\alpha+\delta\ann{a}$, and $\ann{b}=\beta+\delta\ann{b}$,
where $\alpha,\beta\in\mathbb{C}$ denote semiclassical solutions, while $\delta\ann{a},\delta\ann{b}$ describe quantum fluctuations (or Bogoliubov excitations). Semiclassical solutions, also called semiclassical equilibrium points, are found by requiring that all terms linear in the fluctuations vanish in the Liouvillian Eq.~\eqref{Lindblad_ab},
\begin{equation}\label{e. semiclassical equation}
    \begin{dcases}
        (\omega-i\,\gamma)\,\alpha+\lambda\alpha^*+2\epsilon|\alpha|^2\alpha=0~,\\
        (\omega-i\,\gamma)\,\beta+\lambda\beta^*+2\epsilon|\beta|^2\beta=0~.
    \end{dcases}
\end{equation}
Notice that neither the nonlocal bath correlations $m_r$ nor the effective thermal excitations $n_r$ affect the semiclassical equilibrium points.
For a given equilibrium point, the Gaussian approximation is obtained by expanding the Liouvillian up to quadratic order in the fluctuation operators $\delta\ann{a},\delta\ann{b}$. The resulting steady state is Gaussian~\cite{SerafiniBOOK2017quantum} and is therefore fully characterized by the covariance matrix
$\Sigma_{ij}
    =
    \frac{1}{2}\langle \{\hat{z}_i,\hat{z}_j\}\rangle
    -
    \langle \hat{z}_i\rangle\langle \hat{z}_j\rangle ,
$
where $\hat{z}_i\in\{\delta\hat{x}_a,\delta\hat{p}_a,\delta\hat{x}_b,\delta\hat{p}_b\}$ and
$\delta\ann{a}=(\delta\hat{x}_a+i\delta\hat{p}_a)/\sqrt{2}$, with similar definitions for the mode $\hat{b}$. The covariance matrix follows from the Fokker--Planck equation~\cite{BreuerBookOpen, Carmichael_BOOK_1} yielding 
\begin{equation}\label{e. cov matrix equation}
\partial_t\Sigma=F\,\Sigma+\Sigma\,F^T+2\gamma G~.
\end{equation}
The dynamical matrix $F$ depends on the quadratic Hamiltonian around the chosen equilibrium point, whereas the effective thermal noise $n_r$ and nonlocal bath fluctuations $m_r$ enter through the diffusion matrix $G$
\begin{equation}\label{e. G}
    G=\left(n_r +\frac{1}{2}\right)\mathbb{I}_4 - m_r \,\sigma_x\otimes \sigma_z~.
\end{equation}
Since $G$ appears as an inhomogeneous term in Eq.~\eqref{e. cov matrix equation}, the stability of a Gaussian steady state is fully determined by $F$.Therefore, $n_r$ and $m_r$ modify the steady-state Gaussian fluctuations but do not shift the stability threshold.

\begin{figure}[t]  
    \hspace*{-5mm}\includegraphics[width=1\linewidth]{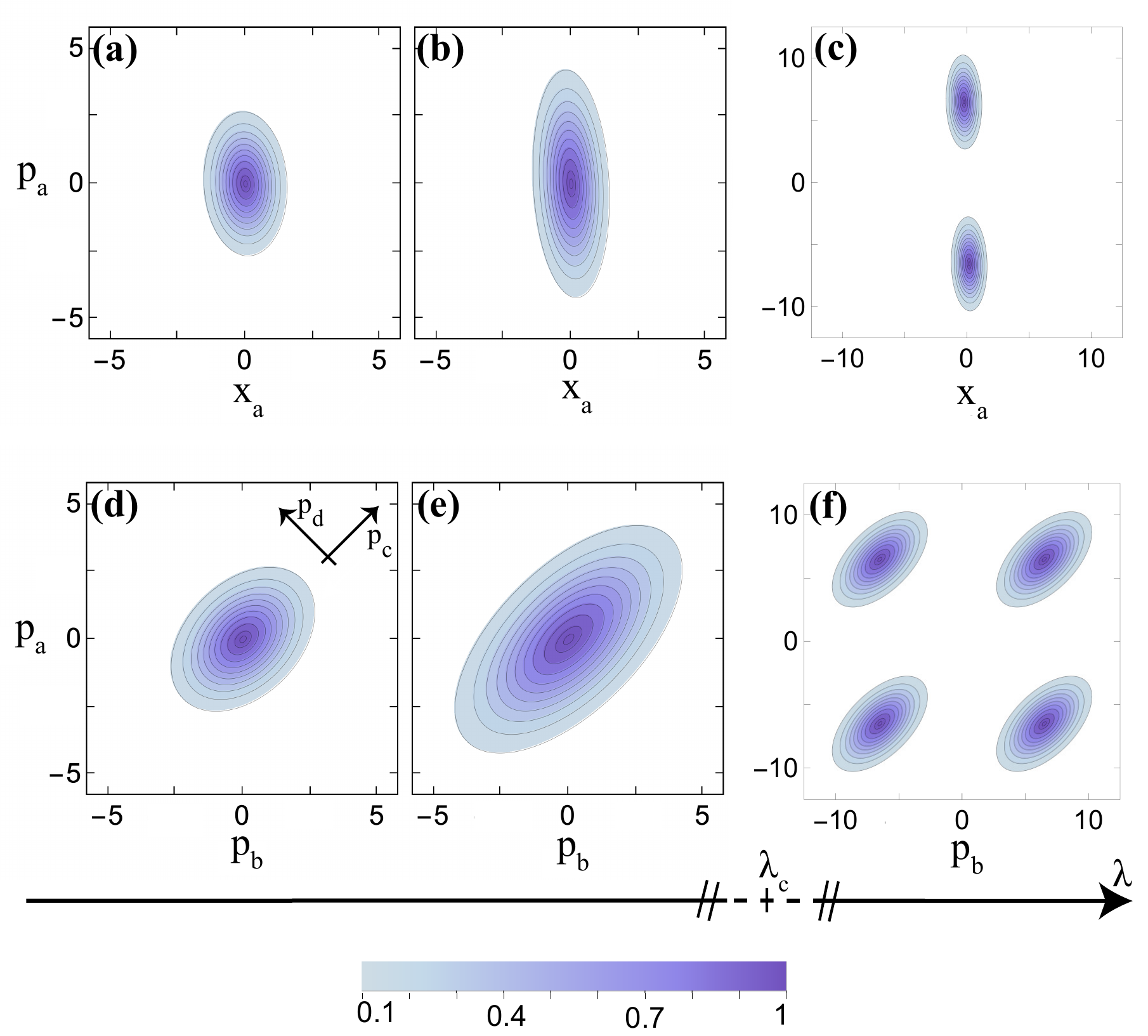}
    \caption{\textbf{Gaussian theory.} Steady-state marginal Wigner functions for various driving amplitudes $\lambda$.
    Panels (a-c): transition as witnessed locally in the $x_ap_a$ plane, which reveals the standard parity-symmetry breaking of parametrically driven Kerr resonators. Panels (d-f): transition as witnessed by joint measurements. Nonlocal bath fluctuations $m_r$ induce correlations in the $p_bp_a$ plane but, under Gaussian approximation, do not modify semiclassical equilibrium points. Parameter values are $\omega=10$, $\gamma=1$, $\epsilon=0.125$, $r=0.5$ and $\lambda=0.5\lambda_c$ (panels $a$ and $d$),  $\lambda=0.8 \lambda_c$ (panels $b$ and $e$), $\lambda=1.5 \lambda_c$  (panel $c$ and $f$). All functions are re-normalized to their maximum.}
    \label{fig:Gaussian}
\end{figure}

Let us first consider the solution to Eq.~\eqref{e. semiclassical equation} continuously connected to the vacuum, that is $\alpha=\beta=0$.
The corresponding drift matrix is
\begin{equation}\label{eq.Fnormal}
    F=    \mathbb{I}_2\otimes\left(i\omega\sigma_y-\lambda\,\sigma_x\right)
    -\gamma\,\mathbb{I}_4~.
\end{equation}
This solution is stable when both the eigenvalues $\mu_\pm = -\gamma \pm \sqrt{\lambda^2-\omega^2}$ of $F$ have negative real part. 
This stability condition gives the critical drive strength $\lambda_c=\sqrt{\omega^2+\gamma^2}$ such that the Gaussian steady state centered at $\alpha=\beta=0$ is stable only for $\lambda<\lambda_c$. This solution corresponds to the normal phase where the parity symmetry is preserved and the order parameter vanishes. Let us analyze the properties of this steady state in the normal phase. Solving Eq.~\eqref{e. cov matrix equation} gives
\begin{equation}
\langle \crea{a}\ann{a}\rangle
=
\langle \crea{b}\ann{b}\rangle
=
n_r+
\left(n_r+\frac{1}{2}\right)
\frac{\lambda^2}{\lambda_c^2-\lambda^2}.
\end{equation}
The first term is the effective thermal contribution inherited from the bath, which remains finite even for $\lambda=0$ [Fig.~\ref{fig:Gaussian}(a)]. The second term is generated by the local parametric drive and diverges as $\lambda\to\lambda_c^-$. Locally, the steady state corresponds to a squeezed thermal state, with a squeezing parameter that increases with $\lambda$ and diverges at the critical point [Fig.~\ref{fig:Gaussian}(b)]. We are interested in the regime where local squeezing amplifies the nonlocal reservoir fluctuations $m_r$. In particular, we choose $\omega\gg\gamma$, so that the local squeezing is approximately aligned along the $\hat{p}_{a}$ and $\hat{p}_b$ quadratures (see Fig.~\ref{fig:Gaussian}(a)-(b) and SI), and we set the squeezing phase to $\phi=0$, such that $m_r$ is real. The nonlocal bath fluctuations  induce correlations both in the  $x_ax_b$ and $p_bp_a$ planes. For the chosen phase relation, the local parametric drive amplifies the correlations in the  $p_bp_b$ plane while suppressing those in the $x_ax_b$ plane. This behavior, progressively enhanced as $\lambda$ approaches $\lambda_c$, is illustrated by the  marginal Wigner distributions in Figs.~\ref{fig:Gaussian}(d) and (e).
\label{s. Gaussian regime}

Let us now analyze the symmetry-broken phase. For $\lambda>\lambda_c$, the solution $\alpha=\beta=0$ becomes unstable. Stable solutions of Eq.~\eqref{e. semiclassical equation} are given by $(\alpha,\beta)=|g|e^{i\varphi}(\pm1,\pm1)$ with independent signs for the two modes, where
\begin{equation}\label{e. semiclassical solutions}
\begin{dcases}
    2\epsilon|g|^2=\sqrt{\lambda^2-\gamma^2}-\omega~,\\
    \lambda\sin(2\varphi)=\gamma .
\end{dcases}
\end{equation}
These four solutions are therefore degenerate and become macroscopically separated in phase space in the limit $\epsilon\to0$, since $|g|^2\sim \epsilon^{-1}$. The system thus enters a symmetry-broken regime in which parity is  spontaneously broken, as shown in Figs.~\ref{fig:Gaussian}(c) and (f).

The Gaussian theory above threshold is obtained by expanding the Liouvillian to quadratic order around each of these four equilibrium points. The dissipative part retains the same form as in Eq.~\eqref{Lindblad_ab} and the quadratic Hamiltonian reads
\begin{align}\label{e. H2 above}
\hat{H}
&=
\Omega\,
\left(
\delta\crea{a}\delta\ann{a}
+
\delta\crea{b}\delta\ann{b}
\right)
+
\frac{\Lambda}{2}
\left(
\delta\hat{a}^2
+
\delta\hat{b}^2
\right)
+
h.c. ,
\end{align}
where
\begin{equation}\label{e. Omega, Lambda}
\begin{dcases}
\Omega=2\sqrt{\lambda^2-\gamma^2}-\omega~,\\
\Lambda=\lambda_c\,e^{i\theta}~,\\
\lambda\sin\theta=\gamma\left(\omega-\sqrt{\lambda^2-\gamma^2}\right)~.
\end{dcases}
\end{equation}
For $\lambda>\lambda_c$, the corresponding dynamical matrix is indeed stable for all four symmetry-broken equilibrium points  (see SI).  
The Gaussian steady state in the symmetry-broken phase can  be formally written as the classical mixture~\cite{Minganti2018}
\begin{equation}
    \rho=
    \frac{1}{4}
    \left(
    \rho_{--}
    +
    \rho_{-+}
    +
    \rho_{+-}
    +
    \rho_{++}
    \right),
\end{equation}
where $\rho_{\pm\pm}$ is the Gaussian steady state of the  fluctuation around the equilibrium point
$(\alpha,\beta)=g(\pm1,\pm1)$. The corresponding Wigner function is depicted in Figs.~\ref{fig:Gaussian}(c) and (f).

Note that within the Gaussian theory, the nonlocal fluctuations $m_r$ of the environmental modes contribute to the fluctuations of the system around each semiclassical solution, but they do not qualitatively alter the structure of the symmetry breaking: the four equilibrium points remain dictated by the local semiclassical equations.
However, Fig.~\ref{fig:Gaussian} suggests that this conclusion may be incomplete near criticality.
The transition between panels (e) and (f) reveals that fluctuations preferentially develop along a specific direction in the $p_b p_a$ plane, yet this anisotropy is not reflected in the symmetry breaking. In particular, one would expect the emergence of lobes aligned with the squeezing direction, similarly to what is observed in panels (b–c). As the Gaussian theory is valid only sufficiently far from $\lambda_c$, it is reasonable to expect a qualitatively different behavior in the critical region. In the following, we will solve this inconsistency of the Gaussian theory with both numerical and analytical methods.

\section{Full-quantum numerical simulations}

\begin{figure*}[h   t]
    \centering
    \includegraphics[width=1\linewidth]{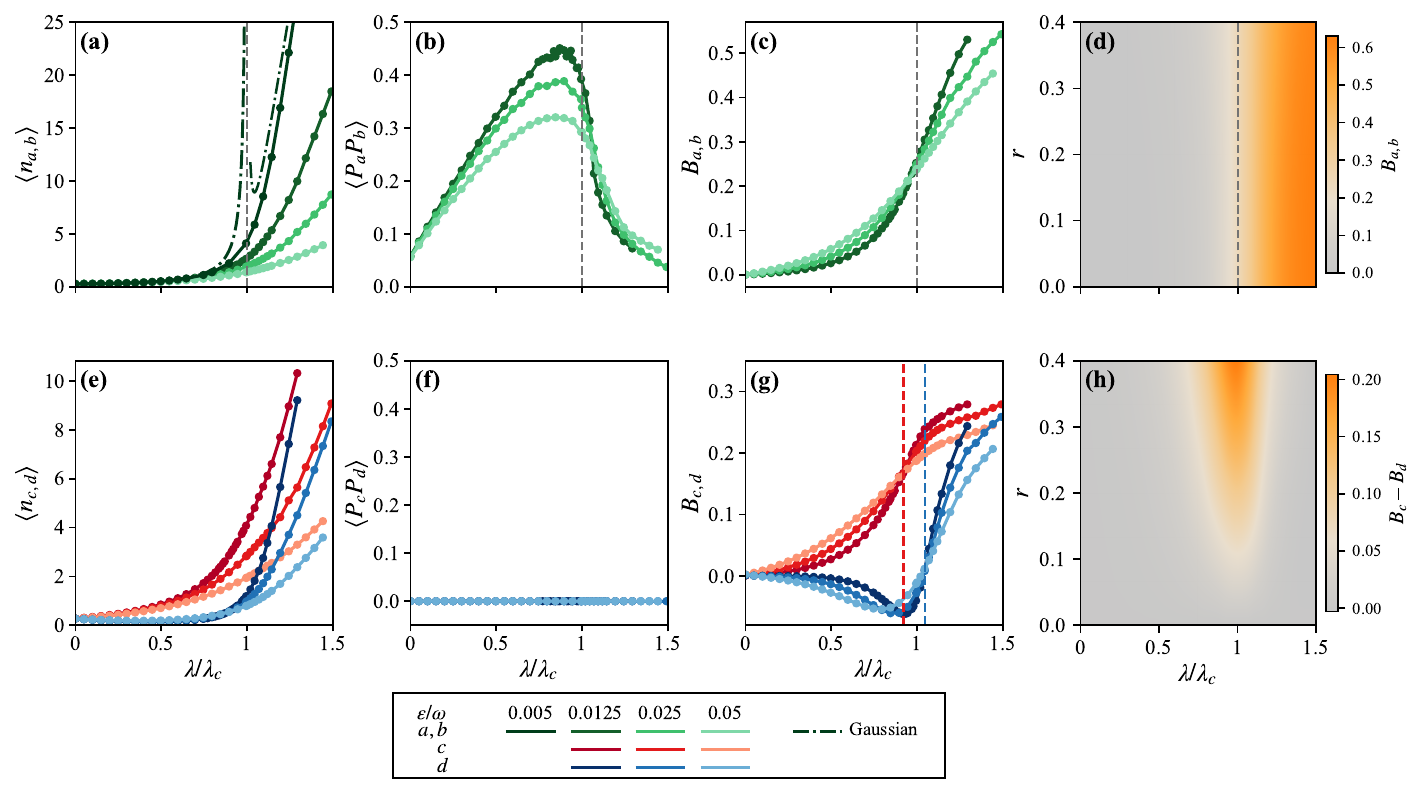}
    \caption{\textbf{Full-quantum numerical simulations.} Results as a function of the drive strength $\lambda$, for $\gamma=0.1\omega$, $\epsilon = \{0.0125, 0.025, 0.05\}\omega $ and $r = 0.5$. Panels (a-d): transition described in terms of the local modes $\ann{a}$ and $\ann{b}$. (a) Photon number shows an increasingly steep transition for decreasing $\epsilon$. As expected, the Gaussian theory is valid for small $\epsilon$ and far from the critical point. (b) Correlations between $\hat{p}_a$ and $\hat{p}_b$ grow in the normal phase and vanish when the critical point is crossed. (c,d) The Binder cumulant shows the non-Gaussian nature of the steady state above the critical point, due to the symmetry breaking. Panels (e-h): transition described in terms of the nonlocal collective modes $\ann{c}$ and $\ann{d}$. (e) Photon numbers reveal the existence of different critical points for $\ann{c}$ and $\ann{d}$. (g) Binder cumulants of collective modes confirm the asymmetric behavior, signature of nonlocal symmetry breaking. Different critical points, marked by vertical dashed red/blue lines, can be identified by the scaling of the Binder cumulant for different $\epsilon$. (h) The difference between the Binder cumulants for $\ann{c}$ and $\ann{d}$ shows that the size of the region of nonlocal symmetry breaking grows for larger nonlocal bath fluctuations $m_r$.}
    \label{fig:numerics}
\end{figure*}

To analyze the behavior of the system across the critical region, we perform exact numerical simulations of the full quantum model. This approach does not require approximations, but is limited to finite values of $\epsilon$ (see Methods).
 Figure~\ref{fig:numerics}(a-d) shows the critical behavior of the $a$ and $b$ modes. Decreasing $\epsilon$ pushes the system toward the thermodynamic limit $\epsilon \to 0$. Panel~(a) shows the transition from vacuum to a highly populated phase, with the vertical dashed line indicating the critical point as predicted by the Gaussian theory, which is accurate well away from criticality and improves as $\epsilon \to 0$. Panel~(b) confirms the expectation suggested by Fig.~\ref{fig:Gaussian}(f): increasing $\lambda$ builds up nonlocal correlations $\langle \hat{p}_a \hat{p}_b\rangle$, which collapse at the critical point.
 
 To investigate the emergence of a symmetry-broken phase, we calculate the Binder cumulant $B_i = 1-\langle \hat{p}_i^4 \rangle/3\langle \hat{p}_i^2 \rangle^2$ of the steady state for $i=a,b$. This is a standard measure of non-Gaussianity in probability distributions~\cite{Binder1981}. In particular, any finite value of $B_i$ is a witness of non-Gaussianity. In panel (c), the Binder cumulant indicates the transition from a Gaussian distribution (vacuum or squeezed-vacuum state) to 
 a bimodal distribution, typical of symmetry-broken states. Local observables and Binder cumulant are consistent with the intuition given by Fig.~\ref{fig:Gaussian} (b) and (c) and reflect the standard phenomenology of parametrically driven Kerr resonators~\cite{Minganti2018,di2023critical,beaulieu_observation_2025}, with the addition of an effective thermal environment populated by $n_r$ photons.

 To unveil the nonlocal nature of the phase transition, we focus on the collective modes $\ann{c} = (\ann{a}+\ann{b})/\sqrt{2}$ and $\ann{d} = (\ann{a}-\ann{b})/\sqrt{2}$, corresponding to a $45^\circ$ rotation of the quadratures, as illustrated in the inset of Fig.~\ref{fig:Gaussian}(d). These variables are defined to decouple the environmental correlations (see Methods): in terms of $\hat{c}$ and $\hat{d}$, the original two-mode squeezing in the bath modes is turned into local single-mode squeezing, with opposite phases for the baths of $\hat{c}$ and $\hat{d}$, respectively. We show the critical behavior of these collective modes in Fig~\ref{fig:numerics} (e-h). In panel (e), the photon numbers  $\langle\crea{c}\ann{c}\rangle$ (blue) and $\langle\crea{d}\ann{d}\rangle$ (red) show a transition from vacuum to highly populated states, but the curves are shifted with respect to each other, hinting at different critical points for the two modes. It is important to notice that  \emph{correlations} in the modes $\hat{a},\hat{b}$ yield an \emph{asymmetry} in the critical behavior of the  collective modes $\hat{c},\hat{d}$ (see Methods). Due to the permutation invariance $\hat{a} \leftrightarrow\hat{b}$, the quadratures $\hat{p}_c$ and $\hat{p}_d$ are uncorrelated, as confirmed by Fig.~\ref{fig:numerics}(f). The nonlocal fluctuations $m_r$ lead to an asymmetric behavior of these collective modes. 
 
 The most striking feature is found in Fig.~\ref{fig:numerics} (g) and (h). There, non-local fluctuations $m_r$ lead to the emergence of a regime where the Binder cumulants $B_c$  and $B_d$ display a significantly different behavior: $B_c$ shows that the mode $\hat{c}$ selectively enters a symmetry-broken phase (see Methods). 
 Vertical dashed lines in Fig.~\ref{fig:numerics}(g) indicate the critical point for modes $\hat{c}$ (red) and $\hat{d}$ (blue), defined as the crossing of the Binder cumulants for decreasing 
values of $\epsilon$. These numerical results strongly point to the emergence of a nonlocal symmetry breaking, which can be fully characterized in terms of collective modes delocalized on remote systems. In panel (h), we show the effect of non-local fluctuations on the correlated phase transition. As can be seen by increasing the squeezing parameter $r$, and hence the non-local bath fluctuations $m_r$, the region of asymmetric phase transitions for collective modes increases. Notably, the emergence of this \emph{nonlocal} critical behavior is not grasped by the Gaussian theory. In the following, we develop an analytical theory that can predict this behavior.

\section{Analytical results on critical behavior}\label{s. slaving}
We develop now an approach based on the slaving principle~\cite{hutt_synergetics_2020, dykman2012fluctuating}. We outline here the main steps; all relevant information can be found in Methods, and the full detailed derivation is provided in the SI. Notice first that the eigenvalues $\mu_\pm$ of the dynamical matrix $F$, defined in Eq.~\eqref{eq.Fnormal}, show a separation of timescales when $\lambda\to\lambda_c$. As the critical point is approached, $\mu_+$ vanishes, resulting in the critical slowing down, while $\mu_- \to -2 \gamma$ remains finite. We define then the variables $u$ and $v$ that diagonalize $F$,
\begin{equation}\label{e. u,v def}
\begin{dcases}
u = -x \cos\chi + p \sin\chi\\
v = \phantom{-}x \cos\chi + p \sin\chi
\end{dcases}~,
\qquad
\tan\chi = \sqrt{\frac{\lambda - \omega}{\lambda + \omega}}~.
\end{equation}
The variable $v$ is associated with $\mu_-$ and can then be regarded as a \textit{fast} variable, while the variable $u$ is critically \textit{slow} as it is associated with the vanishing eigenvalue $\mu_+$. 
We use the collective modes $\ann{c}$ and $\ann{d}$ introduced in the previous section and represent the density operator in the Wigner–Weyl phase space using the corresponding variables $u_j$ and $v_j$, as $W(u_c,v_c,u_d,v_d,t)$.  The master equation \eqref{Lindblad_a} is mapped onto a Fokker-Planck equation. By considering the regime  $\omega/\gamma\gg 1$, where $u,v$ become commuting variables~\cite{dykman2012fluctuating}, and by expanding at first order in $\epsilon$ and $v_j$, we find
\begin{align}\label{e. Fokker u}
    \partial_{t}W=&\sum_{j\in\{c,d\}}\Big[-\partial_{v_j}(F_{v_j}W)-\partial_{u_j}(F_{u_j}W)\,+\nonumber\\
    &\qquad\;\;+\frac{Q_j}{2}\left(\partial_{v_j}^2+\partial_{u_j}^2  -2 \,\partial_{v_j}\partial_{u_j}\right)W\Big]~,
\end{align}
where the drift terms are 
\begin{equation}\label{e.FuFv}
    \begin{dcases}
        F_{v_j}=\left[\beta\,(u_j^2+3 u_{\overline{j}}^2)-\zeta\right]u_j-2\gamma \,v_j~,\\
        F_{u_j}=-\left[\beta\,(u_j^2+3 u_{\overline{j}}^2)-\alpha\right]u_j~.
    \end{dcases}
\end{equation}
For brevity, we denote by $\bar j $ the index exchange $c\leftrightarrow d$. We defined the parameters $\alpha=\omega(\lambda-\lambda_c+2\epsilon)/\gamma$, which provides a non-Gaussian correction to the critical point, and $\beta=\epsilon\,\omega^3/(2\gamma^3)$, $\zeta=2\epsilon \omega/\gamma$, which are vanishingly small in the limit $\epsilon\rightarrow 0$.  The two-mode squeezing of the environmental modes plays a determinant role in the diffusion coefficients, which are asymmetric for the modes $c$ and $d$,
\begin{equation}\label{e. QuQv}
\begin{dcases}
   Q_c = \gamma e^{2 r}~,\\
   Q_d = \gamma e^{-2 r}~.
\end{dcases}
\end{equation}
The separation of timescales is manifest in the drift terms of Eq.~\eqref{e.FuFv}, where $\partial_{v_j} F_{v_j}\sim 2 \gamma \gg \partial_{u_j} F_{u_j}\sim\epsilon$. 
We can then perform an adiabatic approximation and rewrite the Wigner function as
\begin{equation}\label{e. W slaving}
    W=h_c(v_c|u_c,u_d) \;h_d(v_d|u_c,u_d)\; K(u_c, u_d,t)~,
\end{equation}
with the normalizations conditions $\int\!dv_{j}\, h_j(v_j|u_c,u_d)=\int\! du_{c}du_{d}\,K(u_c, u_d,t)=1$. The functions $h_j(v_j|u_c,u_d)$ define the conditional probability of $v_j$ given $u_c$ and $u_d$ that, within the adiabatic elimination,  are time-independent distributions characterizing the local equilibrium of the fast variables. They can be obtained by imposing vanishing probability current in Eq.~\eqref{e. Fokker u} along the $v_j$-directions. This corresponds to imposing $ 2 F_{\nu_j} W - Q_j \partial_{v_j} W =0  $, while treating $u_j$ as a static external parameters. 
As shown in Methods, $h_j(v_j|u_c,u_d)$ are sharply localized Gaussian distributions, adiabatically adjusted according to the evolution of the slow variable. From the perspective of $u_j$, the variables $v_j$ behave as fluctuating noise sources whose mean effect is captured by the average of the force $F_{u_j}$ over $h_j(v_j|u_c,u_d)$. We can now focus on the steady state of the reduced marginal distribution $K(u_c,u_d,t)$, which captures the relevant slow dynamics. Up to first order in $\epsilon$, this distribution is governed by the reduced two-dimensional Fokker-Planck equation
\begin{equation}\label{e.Fokker_K}
\partial_tK(u_c,u_d,t)=\sum_{j\in\{c,d\}}\partial_{u_{j}}\left(K\;\partial_{u_{j}}U+ \frac{1}{2}\,Q_j\;\partial_{u_{j}}K \right)~,
\end{equation}
with the potential $U$ reading
\begin{align}\label{e.potential}
    U(u_{c},u_{d})=-\frac{\alpha}{2}\left(u^2_{c}+u^2_{d}\right)+\frac{\beta}{4}\left(u^4_{c}+6\,u^2_{c}u^2_{d}+u^4_{d}\right)~.
\end{align}
The sign of the coefficient $\alpha$ depends on $\lambda$, while $\beta$ stays strictly positive, yielding a bifurcation at $\alpha = 0$. Crucially, for small but finite nonlinearity $\epsilon$, $\beta\neq 0$ and the quartic term guarantees that the potential is always bounded from below, so this theory correctly predicts a stable steady state across the entire parameter space.
For $\alpha < 0$, that is, below the driving threshold $\lambda < \lambda_c-2\epsilon$, the potential $U(u_{c},u_{d})$ is a single well centered at the origin. For  $\alpha > 0$, that is, for $\lambda > \lambda_c-2\epsilon$, the potential develops a set of four symmetric wells, corresponding to the four possible symmetry-broken solutions.
So far, we have generalized the approach developed in~\cite{dykman2012fluctuating} to the multi-mode case. This provides an analytical solution, valid when there is no environmental two-mode squeezing $r=0$. In this case, the diffusion coefficients are symmetric $Q_c =Q_d = \gamma$ and the steady state of Eq.~\eqref{e.potential} is given by the Gibbs solution $K_{\rm ss}(u_c,u_d)\propto \exp\curly{-2 U(u_{c},u_{d})/\gamma}$. When $r \neq 0$, the diffusion coefficients become inhomogeneous, and Eq.~\eqref{e.Fokker_K} has yet no direct analytical steady-state solution.

\begin{figure}[t!]
  \hspace*{-2mm}\includegraphics[width=1\linewidth]{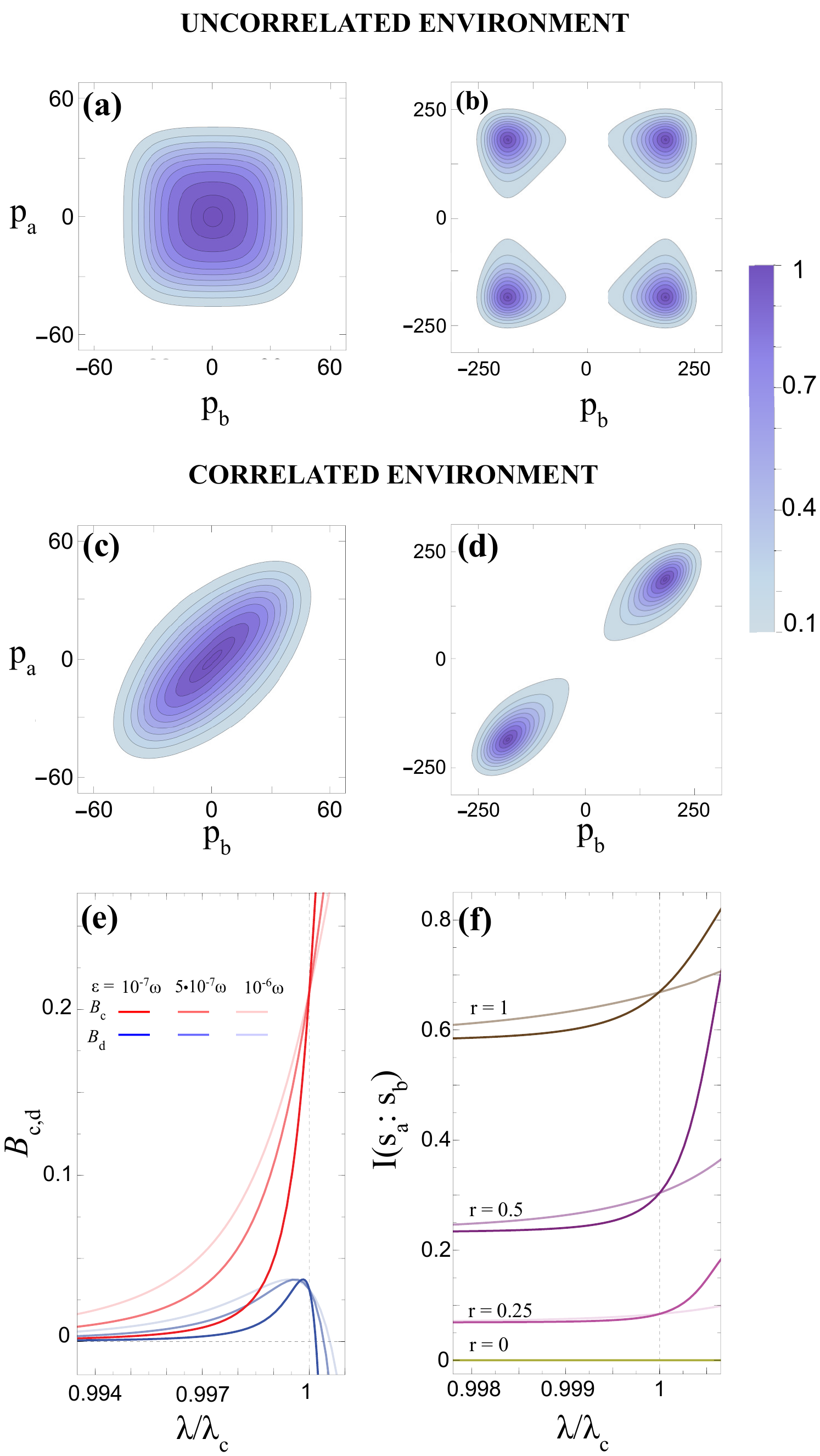} 
  \caption{\textbf{Non-gaussian theory.} (a-d) Steady-state Wigner function in the $p_bp_a$ plane obtained analytically via the slaving principle. In panels (a) and (b) the environments are uncorrelated, $r=0$, whereas, in panels (c) and (d), they are entangled, $r=0.25$. Numerical values are $\omega=10$, $\gamma=1$, $\epsilon=10^{-5}$,  and $\lambda=0.9998\lambda_c$ (panels (a) and (c)), $\lambda=1.0018\lambda_c$ (panels (b) and (d)). (e) Binder cumulants of the modes $\ann{c}$ and $\ann{d}$ for a correlated environment and $\lambda\sim\lambda_c$. They are obtained via numerical integration of the analytical solution for decreasing values of $\epsilon=\left\{1,0.5,0.1\right\} 10^{-6}\omega$. Other numerical values are $\gamma=0.1\omega$ and $r=0.25$. (f) Mutual information shared between the coarse-grained variables $s_a$ and $s_b$ for increasing bath correlations. Lighter lines are for $\epsilon=10^{-6}$, darker ones for $\epsilon=10^{-7}$. As in panel (e), we numerically integrated the analytical solution taking $\gamma=0.1\omega$.}
    \label{fig:Slaving}
\end{figure}

In what follows, we show how this theory can be further extended to include the environmental two-mode squeezing. When $r>1$, the diffusion coefficient $Q_c$ becomes exponentially larger than $Q_d$, introducing an additional separation of timescales. Note that this separation originates from the stochastic component of the Fokker--Planck equation. It therefore differs from the separation between the variables $u$ and $v$ which arises from the deterministic drift terms. For large $r$ we can then apply the slaving principle again and define
\begin{equation}\label{e. K decomposition}
    K(u_c,u_d,t)= k(u_c|u_d)\,y(u_d,t)
\end{equation}
with $\int\!du_{c}\, k(u_c|u_d)=\int\! du_{d}\,y(u_d,t)=1$. The distribution $k(u_c|u_d)$ defines the conditional probability of $u_c$ given $u_d$, and it is obtained as a Gibbs distribution after imposing vanishing probability current in Eq.~\eqref{e.Fokker_K} along the $u_c$-direction, that is  
$ 2 k\ \partial_{u_c} U   + Q_c \partial_{u_c}k =0$,  taking $u_d$ fixed. We find
\begin{equation}\label{e. conditional k Gibbs}
    k(u_c|u_d)=\frac{1}{\mathcal{N}(u_d)}e^{-\tfrac{2}{Q_c}U(u_c,u_d)}~,
\end{equation}
with the non-trivial normalization factor
\begin{equation}\label{e. integral N}
   \mathcal{N}(u_d)=\int_{-\infty}^{+\infty} du_c\,e^{-\tfrac{2}{Q_c}U(u_c,u_d)}~.
\end{equation}
From the perspective of the slow variable $u_d$, the fast variable $u_c$ behaves as a fluctuating noise source. 
Accordingly, the marginal steady-state distribution $y$ can be evaluated as $2 \langle F_{u_d}\rangle y - Q_d \partial_{u_d} y = 0$,
where the average drift reads
\begin{equation}\label{e. average drift Fud}
   \langle F_{u_d}\rangle=\int_{-\infty}^{+\infty}d u_c\,k\,F_{u_d}=\frac{Q_c}{2}\,\,\partial_{u_d}\log\mathcal{N}(u_d)~.
\end{equation}
Up to a normalization factor, we finally obtain a full analytical solution  $y \propto[\mathcal{N}(u_d)]^{\,Q_c/Q_d}$, whose full expression can be formally written in terms of Hermite functions and is provided in Methods. This analytical solution is one of the main results of this work, whose properties are plotted in Fig.~\ref{fig:Slaving}.

First, we notice that it is exact for $Q_c = Q_d$ and in the $Q_c\gg Q_d$ limit, corresponding to $r=0$ and $e^{4r}\gg1$, respectively. However, in the SI, we show that this solution also provides a good approximation for intermediate values of $r$. To obtain a readable solution, let us consider an approximate form of the integral \eqref{e. integral N},
\begin{equation}\label{e. Kss ucud}
    K_{ss}(u_c,u_d)\propto \;e^{-\tfrac{2 U(u_c,u_d)}{Q_c}-\tfrac{u_d^2}{2\sigma}  }
\end{equation}
with $\sigma=\sqrt{\gamma^4/(4\omega^3\epsilon)}\ e^{r}(e^{4r}-1)^{-1}$. This simplified form is written as a Gibbs distribution induced by the potential $U$, supplemented by a delta-like filter that, for $r>1$, is sharply localized along the $u_c$-direction. 
This confinement inhibits the critical behavior of the mode $\ann{d}$ preventing the symmetry breaking in the corresponding phase-space direction. As a result, the phase transition takes place selectively for mode $\ann{c}$, consistently with the numerical data of Fig.~\eqref{fig:numerics}(h).

We can finally go back to the original coordinates $(x_a,p_a,x_b,p_b)$ and write the marginal distribution in the $p_bp_a$ plane as
\begin{equation}\label{e. Wsspapb}
    W_{ss}(p_a,p_b)=\frac{1}{Z}\,e^{-\tfrac{U_{\rm{eff}}(2\chi p_a,2\chi p_b)}{Q_c}}
\end{equation}
with
\begin{equation}\label{e. Ueff}
    U_{\rm{eff}}(p_a,p_b)=-\alpha(p_a^2+p_b^2)+\beta(p_a^4+p_b^4)+\xi (p_a-p_b)^2
\end{equation}
and $2\xi=e^{r}(e^{4r}-1) \sqrt{\omega^3 \epsilon/\gamma^2}$. 
This function offers deeper insight and a robust analytical validation of the phenomenology revealed by our numerical results, operating in a complementary parameter regime. As shown in Fig.~\ref{fig:Slaving}(a,b), when there are no bath correlations ($r=0$), the solution correctly predicts independent symmetry breaking, in agreement with previous theoretical~\cite{Minganti2018} and experimental results~\cite{beaulieu_observation_2025}. For large bath correlations ($r>1$), the analytical solution selectively populates only two minima of the effective potential, corresponding to a correlated symmetry breaking: the signs of the macroscopic displacements acquired by the two resonator modes are locally random yet mutually locked, as shown in Fig.~\ref{fig:Slaving}(d). This behavior is also reflected in the Binder cumulant shown in Fig.~\ref{fig:Slaving}(e), which is in line with the numerical prediction of Fig.~\ref{fig:numerics}(g) in proximity of the critical point. Notably, the analytical solution \eqref{e. Wsspapb} describes the system behavior in the effective thermodynamical limit $\epsilon\to 0$ where numerical simulations are subject to strong limitations. Finally, the nonlocal character of the symmetry breaking is clearly illustrated in Fig.~\ref{fig:Slaving}(f), which shows that for any finite environmental correlation $r \neq 0$, the mutual information $I(s_a\,{:}\,s_b)$ between the coarse-grained variables $s_a=\mathrm{sgn}(p_a)$ and $s_b=\mathrm{sgn}(p_b)$ exhibits a discontinuous jump at $\lambda=\lambda_c$. This highlights that the steady state of the two modes $\ann{a}$ and $\ann{b}$ becomes strongly correlated as the critical point is crossed (see Methods). Overall, this analytical derivation clarifies the key mechanism behind the emergence of a nonlocal critical mode. The nonlocal correlations of the bath lead to asymmetric diffusion coefficients for the collective modes $\hat{c}$ and $\hat{d}$, thereby enhancing the critical behavior of the former while suppressing that of the latter.

\section{Discussion}
The principal significance of these results is the introduction of a conceptual framework for critical phenomena, in which nonlocal quantum fluctuations emerge as fundamental mechanism driving phase transitions. 
A key physical implication is that nonlocal environmental correlations can lead to the emergence of \emph{collective} critical modes that are delocalized over spatially remote systems. The phase transition can only be fully characterized through joint nonlocal measurements, while a local observer only witnesses a phase transition driven by thermal fluctuations. These findings have been obtained via both numerical and analytical methods, which are valid in complementary parameter regimes. Numerical simulations, feasible only for finite-size systems, show that the fingerprints of nonlocal phase transitions emerge even for relatively large nonlinearities. Analytical results cover the phenomenology in the effective thermodynamic limit and across the critical region. They validate the picture suggested by numerical results and provide a deep physical insight into the predicted phenomenology. Most notably, our analytical derivation explains both the emergence of nonlocal critical modes and the correlated symmetry breaking driven by nonlocal quantum fluctuations.

Our setting is fundamentally distinct from the phenomenology of anomalous heat flows and generalized thermodynamic laws~\cite{del_rio_relative_2016, bera_generalized_2017}, which hinge on system–environment correlations. It is also unrelated to models involving direct coupling between critical systems, such as the Ashkin–Teller and Potts models~\cite{alcaraz_surface_1987}. In our case, the critical systems are not coupled to each other, nor is there a direct interaction between their respective environments.
The considered model can be implemented with current quantum platforms. In the context of superconducting quantum circuits, critical systems can be directly implemented with Kerr resonators~\cite{Chen2023,beaulieu_observation_2025} or even with quantum acoustic modes~\cite{scigliuzzo_quantum_2026}. Notice that two-mode squeezed states have been recently used to entangle superconducting qubits~\cite{juanes_entangling_2025} and that a cryogenic waveguide can be used to connect systems located in separate refrigerators~\cite{magnard_microwave_2020, fedorov_experimental_2021, casariego_propagating_2023, yam_cryogenic_2025}. Most notably, these microwave links have allowed for a loophole-free violation of Bell inequalities~\cite{storz_loophole-free_2023}, demonstrating that genuinely nonlocal properties can be explored in this platform. Vibrational modes of trapped ions~\cite{cai2021observation} can also be used to implement dissipative phase transitions of finite-component models, and they can be provably connected to entangled signals~\cite{krutyanskiy_entanglement_2023}.

The research direction opened by this work goes far beyond the specific model here considered. Cold-atom experiments~\cite{baumann_dicke_2010, nagy_dicke-model_2010, klinder_dynamical_2015, Ferri2021, helson_density-wave_2023, song_dissipation-induced_2025} show that the same framework is within reach of current quantum many-body systems. We derived our results in the context of driven-dissipative phase transitions. However, our framework can be generalized to systems that are at thermal equilibrium with their environment, and even applied to materials strongly coupled to cavity modes~\cite{garcia-vidal_manipulating_2021}. 
In conclusion, our findings call for a systematic theoretical and experimental exploration of nonlocal critical phenomena. Besides the fundamental interest, promising directions include applications in critical quantum sensing~\cite{montenegro_review_2025, mihailescu2025tutorial, stas_entanglement-assisted_2026,alushi2025collective} and distributed quantum information processing~\cite{knorzer2025distributed}.

\section{Methods}
\textbf{Collective modes.}
The modes $\ann{c}=(\ann{a}+\ann{b})/\sqrt{2}$ and $\ann{d}=(\ann{a}-\ann{b})/\sqrt{2}$ are defined such that the environmental correlations are decoupled, and the master equation~\eqref{Lindblad_ab} becomes
\begin{equation}\label{e. master cd Met}
    \partial_t\rho=-i\,[H_{int},\rho]+\mathcal{L}_c[\rho]+\mathcal{L}_d[\rho]~,
\end{equation}
where $\mathcal{L}_c[\rho]=-i [H_c,\rho]+ \mathcal{D}_c^{th}[\rho]+\mathcal{D}_c^{sq}[\rho]$, with
\begin{align}
    \!\!\!\begin{dcases}
        H_c=\omega c^\dagger c+\frac{1}{2}\left(\lambda \,c^2 + \lambda^* c^{\dagger 2}\right)~,\\
        \mathcal{D}_c^{th}[\rho]=\gamma (1+n_r) \left(2c\rho c^\dagger- \{c^\dagger c,\rho\}\right)+\gamma n_r \left(2c^\dagger\rho c- \{c c^\dagger,\rho\}\right)~,\\
        \mathcal{D}_c^{sq}[\rho]=\gamma m_r \left(2c^\dagger\rho c^\dagger- \{c^{\dagger 2},\rho\}\right)+\gamma m_r^* \left(2c\rho c- \{c^2,\rho\}\right)~,
        \end{dcases}
\end{align}
$\mathcal{L}_d[\rho]=\mathcal{L}_c|_{m_r\rightarrow -m_r}[\rho]$ and
\begin{equation}\label{e. Hint Met}
    H_{int}=\frac{\epsilon}{2}\left(c^{\dagger 2}+d^{\dagger 2}\right)\left(c^2+d^2\right)+\epsilon \,c^\dagger c\, d^\dagger d~.
\end{equation}
Note that the new master equation~\eqref{e. master cd Met} does not contain correlated dissipative terms between the collective modes $\hat{c}$ and $\hat{d}$. The modes $\ann{c}$ and $\ann{d}$ are coupled only through the Hamiltonian non-linear interaction in Eq.~\eqref{e. Hint Met}. 

Let us now discuss some key properties of these modes. First, we note that for permutationally invariant $\hat{a} \leftrightarrow\hat{b}$ systems, the quadratures $\hat{p}_c$ and $\hat{p}_d$ are uncorrelated, as shown in Fig.~\ref{fig:Gaussian}(f). This can be proven analytically for any odd power $n=2l+1$ ($l\in\mathbb{N}$) of the quadratures. Indeed, it is
\begin{align}
\langle \hat{p}_c^n \hat{p}_d^n\rangle &\propto 
\langle (\hat{p}_a + \hat{p}_b )^n(\hat{p}_a - \hat{p}_b )^n \rangle=
\langle (\hat{p}_a^2 - \hat{p}_b^2 )^n \rangle = \nonumber\\
&=\langle (\hat{p}_b^2 - \hat{p}_a^2 )^n \rangle=(-1)^n\langle \hat{p}_c^n \hat{p}_d^n\rangle~,
\end{align}
yielding $\langle \hat{p}_c^n \hat{p}_d^n\rangle=0$.
We show now that, by construction, correlations in the local modes $\ann{a}$ and $\ann{b}$ correspond to asymmetry in observables related to the collective modes $\ann{c}$ and $\ann{d}$. By direct inspection, for the photon number, we find 
$\crea{c}\ann{c} - \crea{d}\ann{d} = \langle \crea{a} \ann{b}\rangle+\langle \ann{a} \crea{b}\rangle$. For the quadratures, it is
$\langle\hat{p}_c^2\rangle - \langle\hat{p}_d^2\rangle = 2\langle \hat{p}_a \hat{p}_b\rangle$, where we used the permutational symmetry $\langle\hat{p}_a\rangle\leftrightarrow\langle\hat{p}_b\rangle$. Finally, considering the definition of the Binder cumulants $B_i = 1-\langle \hat{p}_i^4 \rangle/3\langle \hat{p}_i^2 \rangle^2$, for collective modes it is
\begin{equation}
B_c - B_d
=
\frac{1}{3}
\left[
\frac{\left\langle (p_a - p_b)^4 \right\rangle}
{\left\langle (p_a - p_b)^2 \right\rangle^{2}}
-
\frac{\left\langle (p_a + p_b)^4 \right\rangle}
{\left\langle (p_a + p_b)^2 \right\rangle^{2}}
\right].
\end{equation}
This figure of merit quantifies the directional selectivity of symmetry breaking in phase space. When $B_c-B_d > 0$, the marginal Wigner function on the plane $p_bp_a$ becomes increasingly bimodal in the symmetric variable $p_c = p_a + p_b$, while remaining uni-modal in the antisymmetric one $p_d = p_a - p_b$, as in Fig.~\ref{fig:Slaving}(d). By contrast, when $B_c-B_d = 0$, the non-Gaussianity is equal along the two directions, as in Fig.~\ref{fig:Slaving}(b). This behavior can be understood by noting that the difference $B_c - B_d$ is odd under the local-parity transformation $p_a \rightarrow -p_a$, or $p_b \rightarrow -p_b$. Consequently, if the steady state preserves the symmetry, as occurs when $r = 0$ or for finite $r$ sufficiently far from the critical region, the difference $B_c - B_d$ vanishes.\\

\textbf{Adiabatic elimination.}
As outlined in Sec.~\ref{s. slaving}, the analytical expression \eqref{e. Kss ucud} follows from a two-step adiabatic elimination procedure. The first step relies on the deterministic timescale separation between the fast variables $v_j$ and the critically slow variables $u_j$, implying the factorization \eqref{e. W slaving} for the Wigner function ($j=c,d$). By treating $u_j$ as static parameters, the conditional probabilities $h_j(v_j|u_c, u_d)$ follows from the stationary Fokker-Planck equation $2 F_{v_j} W - Q_{j} \partial_{v_j} W = 0$, with $F_{v_j}$ and $Q_j$ as in Eqs.~\eqref{e.FuFv} and \eqref{e. QuQv}. At first order in $\epsilon$, they are Gaussian distributions
\begin{equation}
    h_j(v_j|u_{c},u_{d}) = \sqrt{\frac{2\gamma}{\pi\, Q_j}} \; e^{-\tfrac{2\gamma}{Q_j} \left[v_{j} - v_{j}^{ad}(u_{c}, u_{d})\right]^2}~,
\end{equation}
that, because $\gamma \gg \epsilon$, are sharply localized around their centers
\begin{equation}
    v_j^{ad}(u_c,u_d) = \left[ \beta \left( u_j^2 + 3 u_{\bar{j}}^2 \right) - \zeta \right] \frac{u_j}{2\gamma}~,
\end{equation}
which define the adiabatically instantaneous equilibrium of the fast variables. Averaging the dynamics over $h_j$ yields the reduced Fokker-Planck Eq. \eqref{e.Fokker_K} for the marginal distribution $K(u_c, u_d, t)$.

A second timescale separation arises from the bath correlations inducing asymmetric diffusion coefficients in Eq.~\eqref{e.Fokker_K}. The variable $u_d$ evolves much more slowly than $u_c$, allowing $K$ to be further factorized as in Eq.~\eqref{e. K decomposition}. The steady-state distribution $y_{ss}(u_d)$ follows by the averaging over the conditional probability $k(u_c|u_d)$, given in Eq.~\eqref{e. conditional k Gibbs} and derived by treating $u_d$ as a static parameter. The distribution $y_{ss}(u_d)$ is found proportional to $[\mathcal{N}(u_d)]^{Q_c/Q_d}$, with $\mathcal{N}(u_d)$ defined in Eq.~\eqref{e. integral N}. Evaluating this normalization factor requires integrating the exponential of the quartic potential $U(u_c, u_d)$ over $u_c$. The integral can be rewritten in terms of negative-index Hermite functions
\begin{equation}
    \mathrm{H}_\nu(z)=\frac{1}{\Gamma(-\nu)}\,\int_{0}^{\infty}ds\,\,e^{-(s^2+2zs)}\,s^{-(\nu+1)}~,
\end{equation}
with $\mathrm{Re}(\nu)<0$ and $\Gamma(-\nu)$ the Euler gamma function. As a result, the steady-state marginal distribution $K_{ss}$ takes the form
\begin{equation}\label{e. Kss_Met}
    K_{ss}(u_c,u_d)=\frac{1}{Z}\,\exp\left[-\frac{2}{Q_c} U(u_c,u_d)\right] \,J(u_d^2)~,
\end{equation}
where $Z$ is a normalization constant, and the modulation function 
\begin{align}\label{e. R def Met}
    J(u_d^2)=\exp\Bigg\{\left(\frac{Q_c}{Q_d}-1\right)\;&\bigg[\frac{1}{Q_c}\left(\alpha\,u_d^2-\frac{\beta}{2}u_d^4\right)+ \nonumber\\
    &+\log{\mathrm{H}_{-1/2}\left(\frac{3\beta u_d^2-\alpha}{\sqrt{2\beta Q_c}}\right)}\,\bigg]\,\Bigg\}
\end{align}
describes the effect of the asymmetry between the diffusion coefficients. Eq.~\eqref{e. R def Met} identifies a symmetric curve sharply peaked at $u_d=0$. Anyway, it may be challenging to interpret due to the presence of the Hermite function. For that reason, while the full expression in Eq.~\eqref{e. R def Met} is maintained to generate the results in Fig.~\ref{fig:Slaving}, a more transparent form of $K_{ss}$ is given in Eq.~\eqref{e. Kss ucud}. It is obtained by expanding the exponent in Eq.~\eqref{e. R def Met} at first order in $u_d^2$, 
\begin{equation}\label{e. R approx Met}
    J(u_d^2)\sim\exp\left[-\,\sqrt{\frac{2\beta}{Q_c}}\left(\frac{Q_c}{Q_d}-1\right)\,f\!\left(\frac{\alpha}{\sqrt{2\beta Q_c}}\right)\,u_d^2\right]
\end{equation}
with
\begin{equation}
    f(w)=\frac{3\, \mathrm{H}_{-3/2}(-w)}{2\, \mathrm{H}_{-1/2}(-w)}-w~,
\end{equation}
and by noting that for $\lambda\sim\lambda_c$, it is $\alpha \sim 0$, implying $f(w) \sim 1$. Eq.~\eqref{e. R approx Met} explicitly shows how, under strong bath correlations $Q_c\gg Q_d$, the modulation $J(u_d^2)$ reduces to a delta-like function, enforcing a sharp localization of the Gibbs distribution $\exp\!{[-2U(u_c,u_d)/Q_c]}$ along the $u_c$-direction.\\

\textbf{Coarse-grained mutual information.}
To quantify the correlations between the two resonators, we evaluated in Fig.~\ref{fig:Slaving}(f) the classical mutual information $I(s_a\,{:}\,s_b)$ emerging from a dichotomic coarse-graining of the $p_bp_a$ plane. The Wigner function $W_{\mathrm{ss}}(p_a, p_b)$ in Eq.~\eqref{e. Wsspapb} is projected onto the variables $s_a=\mathrm{sgn}(p_a)$ and $s_b=\mathrm{sgn}(p_b)$, defining a classical joint probability distribution from which the mutual information $I(s_a\,{:}\,s_b)=S(s_a)-S(s_a|s_b)$ is calculated, where $S(\cdot)$ is the Shannon entropy. Due to the $(\hat{a},\hat{b}) \leftrightarrow -(\hat{a},\hat{b})$ symmetry of the Lindbladian, the distribution $W_{\mathrm{ss}}(p_a, p_b)$ is symmetric with respect to the origin. Consequently, the probabilities for the sign of each $p_i$ ($i=a,b$) are equal, \textit{i.e.} $P(s_i=+1) = P(s_i=-1) = 1/2$, yielding $S(s_i) = 1$. As regards the conditional entropy, it is
\begin{align}
    S(s_a|s_b) &= -\sum_{i,j\in\{a,b\}} P(s_i) P(s_j|s_i) \log_2 P(s_j|s_i) \nonumber \\
    &= -\left[ P \log_2 P + (1-P) \log_2 (1-P) \right]~,
\end{align}
where we introduced the conditional probability $P := P(s_a=1 | s_b=1) = P(s_a=1 \cap s_b=1) / P(s_b=1)$ reading
\begin{align}
    P &= 2 \int_{0}^{+\infty} \! \mathrm{d}p_a \int_{0}^{+\infty} \! \mathrm{d}p_b \, W_{\mathrm{ss}}(p_a, p_b)~.
\end{align}
Exploiting the symmetry of $W_{\mathrm{ss}}(p_a, p_b)$, and given that, near the critical point $\lambda \to \lambda_c$, only the slow variables $u_i = 2p_i\sin\chi$ (see SI) stay relevant, we can rewrite $P = 1/(1+\mu)$, with
\begin{align}\label{e. mu def}
    \mu &:= \frac{\int_{-\infty}^{0} \! \mathrm{d}u_a \int_{0}^{+\infty} \! \mathrm{d}u_b \, K_{\mathrm{ss}}\!\left(\frac{u_a+u_b}{\sqrt{2}}, \frac{u_a-u_b}{\sqrt{2}}\right)}{\int_{0}^{+\infty} \! \mathrm{d}u_a \int_{0}^{+\infty} \! \mathrm{d}u_b \, K_{\mathrm{ss}}\!\left(\frac{u_a+u_b}{\sqrt{2}}, \frac{u_a-u_b}{\sqrt{2}}\right)}
\end{align}
and $K_{\mathrm{ss}}$ as in Eq.~\eqref{e. Kss_Met}. 
The parameter $\mu$ represents the relative weight of the Wigner function in the second quadrant of the $p_bp_a$ plane with respect to the first quadrant. The coarse-grained mutual information therefore reads
\begin{align}
    I(s_a\,{:}\,s_b) &= 1 + \frac{\mu}{1+\mu} \log_2 \mu - \log_2 (1+\mu)~.
\end{align}
For an uncorrelated environment $r=0$, as in Fig.~\ref{fig:Slaving}(a)-(b), the two independent local symmetries $p_a \to -p_a$ and $p_b \to -p_b$ yield $\mu=1$ and zero mutual information. Conversely, as the squeezing strength $r$ increases, the ratio $\mu$ tends to zero, driving the information to its maximum value equal to $1$. Fig.~\ref{fig:Slaving}(f) shows the behaviour of $I(s_a\,{:}\,s_b)$ in the vicinity of the critical point $\lambda \sim \lambda_c$. Just below threshold, as in Fig.~\ref{fig:Slaving}(c), the Wigner function becomes highly squeezed yielding a small $\mu$; consequently, $I(s_a\,{:}\,s_b)$ takes a finite positive value that grows with $r$. Just above threshold, as in Fig.~\ref{fig:Slaving}(d), the support of $W_{\mathrm{ss}}(p_a, p_b)$ sharply splits into two disconnected regions, effectively vanishing in the second and fourth quadrants. This forces $\mu$ to drop abruptly to $0$, driving $I(s_a\,{:}\,s_b)$ to its maximum.\\

\textbf{Numerical methods.}
Numerical simulations were performed using the Quantum Toolbox in Python (\texttt{QuTiP}), employing the master-equation \texttt{mesolve} solver to evaluate the system dynamics. The calculations were carried out in the Fock basis, with the Hilbert space truncated at a finite maximum photon number for each cavity. Numerical convergence was systematically verified with respect to both the photon-number cutoff and the evolution time, until the relevant observables remained unchanged within $10^{-3}$ accuracy. All results were further benchmarked against the single-resonator model obtained by tracing out resonator $b$. This model reproduces all local observables exactly and reduces the Hilbert-space dimension by half, providing a stringent consistency check on the numerical implementation.

\begin{acknowledgments}
\noindent
We acknowledge insightful discussions with Luis A. Correa, Marco Baldovin, Dario Lucente, and Federico Carollo.
S.F. acknowledges financial support from the foundation Compagnia di San Paolo, grant vEIcolo no. 121319. A.C., V.B., and S.F. acknowledge from PNRR MUR project PE0000023-NQSTI, financed by the European Union–Next Generation EU. A.J.S. and H.A. acknowledge the support as part of QuPIDC, an Energy Frontier Research Center, funded by the US Department of Energy (DOE), Office of Science, Basic Energy Sciences (BES), under award number DE-SC0025620, as well as the support by the US Department of Energy (DOE), the Office of Basic Energy Sciences (BES), and the Division of Materials Sciences and Engineering under award number DE-SC0025554. R.D. acknowledges financial support from the Academy of Finland, Grants No. 349199, No. 353832, and No. 368477.

\end{acknowledgments}

\clearpage
\newpage
\appendix
\widetext

\begin{center}
\textbf{SUPPLEMENTAL INFORMATION}
\end{center}

\section{Derivation of the Lindblad Master Equation}
Our starting point is the full Hamiltonian that includes both the systems and the baths degrees of freedom, 
$\hat{H}_{\rm Tot} = \hat{H}_{\rm S}(t) + \hat{H}_{\rm Env} + \hat{V}$.
The joint Hamiltonian $\hat{H}_S$ for the systems $a$ and $b$ is time-dependent as it includes the parametric driving term,
\begin{equation}
\hat{H}_S(t) = \omega_r  (\hat{a}^\dagger \hat{a} + \hat{b}^\dagger \hat{b}) + 
\lambda \cos(\omega_d t) \left( \crea{a}\crea{a} + \ann{a}\ann{a} + \crea{b}\crea{b} + \ann{b}\ann{b}  \right)
+ \epsilon \left( \kerr{a} + \kerr{b}  \right).
\end{equation}
The free Hamiltonian of the environments is composed of the sum of two continua of modes,
$\hat{H}_{\rm Env}= \int d\omega\  \omega \round{\hat{A}_\omega^\dagger \hat{A}_\omega + \hat{B}_\omega^\dagger \hat{B}_\omega} $, while the system-bath interactions are given by,
\begin{equation}
\hat{V} = \int d\omega \frac{g(\omega)}{\sqrt{2\pi}} \left[
\round{\crea{a} + \ann{a}} \left(\hat{A}^\dagger_\omega + \hat{A}_\omega \right) +
\round{\crea{b} + \ann{b}}\left(\hat{B}^\dagger_\omega +\hat{B}_\omega \right)
\right].
\end{equation}
We now switch to a rotating frame with respect to the free Hamiltonian $\hat{H}_{\rm RWA} = \frac{\omega_d}{2}(\crea{a}\ann{a} + \crea{b}\ann{b})+\hat{H}_{\rm Env}$, chosen to make the system Hamiltonian time independent:
\begin{equation}
\label{H_Kerr_full}
\hat{H}^R_S = \Delta (\hat{a}^\dagger \hat{a} + \hat{b}^\dagger \hat{b}) 
+ \frac{\lambda}{2} \left( \crea{a}\crea{a} + \ann{a}\ann{a} + \crea{b}\crea{b} + \ann{b}\ann{b}  \right)
+ \epsilon \left( \kerr{a} + \kerr{b}  \right),
\end{equation}
where we defined the (two-photon) drive-to-resonator detuning $\Delta = \omega_r - \omega_d/2$, and we neglected the fast-oscillating terms of frequency $\omega_d$.
In the rotating frame, the system-bath coupling Hamiltonian reads, 
\begin{equation}
\hat{V}^R(t) = \int d\omega \frac{g(\omega)}{\sqrt{2\pi}} \left[
\round{e^{i(\omega_r-\Delta) t}\crea{a} + e^{-i(\omega_r-\Delta) t}\ann{a}} \left(e^{i\omega t}\hat{A}^\dagger_\omega + e^{-i\omega t}\hat{A}_\omega \right) +
\round{e^{i(\omega_r-\Delta) t}\crea{b} + e^{-i(\omega_r-\Delta) t}\ann{b}}\left(e^{i\omega t}\hat{B}^\dagger_\omega +e^{-i\omega t}\hat{B}_\omega \right)
\right].
\end{equation}
We will now use this rotating frame Hamiltonian to develop a microscopic derivation of the master equation, which is valid under the assumption that the system-bath coupling is weak and that $\omega_r$ is the dominant frequency. We take the drive to be very close to the two-photon resonance condition, that is,  
we consider the energy scale $\epsilon \ll \Delta, \lambda \ll \omega_r,\omega_d $. Notice that this condition is the most natural and less demanding for experimental implementations of the considered driven-dissipative phase transition~\cite{beaulieu_observation_2025,beaulieu_criticality-enhanced_2025}. Indeed, to achieve the critical point, the drive intensity $\lambda$ must be comparable to the detuning $\Delta$, which can be an order of magnitude smaller than the bare resonator frequency $\omega_r$.  Under these conditions, the parametric drive has a dramatic effect on the system steady-state properties,
but it does not qualitatively modify the dissipative kernel \cite{wagner2026}.

Switching to the interaction picture with respect to the system's Hamiltonian $\hat{H}^R_S$, we can write the Liouville equation for the full density operator, including system and environment, as

\begin{equation}
\label{EqMotRho}
\dot{\rho}_{\rm Tot}(t) = - i\squared{\hat{V}_I^R(t),\rho_{\rm Tot}(t)}\ 
\end{equation}
where $\rho_{\rm Tot}(t)$ and $\hat{V}_I^R(t)$ denote respectively the density matrix and the system-bath interaction in the interaction picture, {\sl i.e.} $\hat{V}_I^R(t)=U_S^\dagger(t)\hat{V}^R(t)U_S(t)$ with $U_S(t)=e^{-i\hat{H}^R_S t}$. %
Integrating the above equation between $t$ and $t+\Delta t$, and expanding to the second order in the interaction $\hat{V}$, we obtain
\begin{equation}
\rho(t+\Delta t)-\rho(t)=-\frac{i}{\hbar}\int_t^{t+\Delta t} dt^\prime \,{\rm Tr}_E\left\{[\hat{V}_I^R(t'),\rho_{\rm Tot}(t)]\right\}- \frac{1}{\hbar^2}\int_t^{t+\Delta t} dt^\prime \int_t^{t^{\prime}} \!\!\!d t^{\prime\prime}\, {\rm Tr}_{\rm Env}\left\{[[\hat{V}_I^R(t^{\prime}),[\hat{V}_I^R(t^{\prime\prime}),\rho_{\rm Tot}(t)]\right\}~,\label{ME1}
\end{equation}
with $\rho(t)$ denoting the reduced density operator for the system. Assuming that the bath correlation time, which sets the decay of the corresponding two-time correlation functions, is much shorter than the typical evolution 
timescale of the interaction-picture density matrix, in Eq.~\eqref{ME1}, we can factorize the 
total density matrix as $\rho_{\rm Tot}(t) \simeq \rho_E(t) \otimes \rho(t)$.
Under the assumption ${\rm Tr}_E\{\hat{V}_I^R(t)\,\rho_E(t)\}=0$, 
and expanding the system-bath coupling to the second order, the equation of motion for the system's density matrix reads~\cite{BreuerBookOpen, cohen2024Book}
\begin{equation}
    \dot{\rho}(t) = -\frac{1}{\hbar^2}\int_0^\infty \!\!d\tau\;
    {\rm Tr}_{\rm Env}\!\left[\hat{V}_I^R(t),
    \bigl[\hat{V}_I^R(t-\tau),\,\rho_E(t)\otimes\rho(t)\bigr]\right].
\end{equation}
where we made a change of variable in the integral, and we extended the integration range to infinity.
Switching back to the Schr\"odinger picture, we obtain:
\begin{equation}
\label{master_trace1}
\dot{\rho}(t) = 
 - i \squared{\hat{H}^R_S,\rho(t)} -
\int_0^\infty d\tau \Tr_{\rm Env} \left\{[[\hat{V}_I^R(t),[U_S(t)\hat{V}_I^R(t-\tau)U^\dagger_S(t),\rho_E (t)\otimes\rho(t)]\right\}. 
\end{equation}
We now introduce a further approximation, namely, in the second term in Eq.~\eqref{master_trace1}, we neglect the evolution induced by terms of the order $\Delta$, $\lambda$, and $\epsilon$. This is justified under the assumption that the spectral density is 
approximately constant over these energy scales, which are order of magnitudes smaller than $\omega_r$ \cite{wagner2026}. 
We then eventually arrive at the master equation in its final form:
\begin{equation}
\label{master_trace2}
\dot{\rho}(t) = 
 - i \squared{\hat{H}^R_S,\rho_S(t)} -
\int_0^\infty d\tau \Tr_{\rm Env} \left\{[[\hat{V}^R(t),[\hat{V}^R(t-\tau),\rho_E (t)\otimes\rho(t)]\right\}. 
\end{equation}
To perform the trace over the environment degrees of freedom, it is convenient to rewrite the interaction Hamiltonian as
\begin{equation}
\hat{V}^R= \sum_{\nu\in \curly{+,-} } e^{i \nu \omega_r t}
\round{ \hat{a}_\nu \hat{X}_A(t) + \hat{b}_\nu \hat{X}_B(t) },
\end{equation}
where we relabeled the creation and annihilation operators $\hat{a}_+ = \hat{a}^\dagger$, $\hat{a}_- = \hat{a}$, and similarly for $\hat{b}_\nu$. We defined also the operators
\begin{align}
\hat{X}_A(t) &= \int d\omega \frac{g(\omega)}{\sqrt{2\pi}} 
\round{e^{i\omega t} \hat{A}^\dagger_\omega  + e^{-i\omega t} \hat{A}_\omega  }~,\\
\hat{X}_B(t) &= \int d\omega \frac{g(\omega)}{\sqrt{2\pi}} 
\round{e^{i\omega t} \hat{B}^\dagger_\omega  + e^{-i\omega t} \hat{B}_\omega}~.
\end{align}
The Master equation Eq.~\eqref{master_trace2} can then be rewritten as
\begin{align}
\dot{\rho}(t) &= 
- i \squared{\hat{H}_S^R,\rho(t)} + \nonumber\\ 
&+ \int_0^t d\tau \Bigg\{ \sum_{\nu,\nu^\prime} e^{i \omega_r \round{\nu+\nu^\prime}t} e^{-i \omega_r \nu \tau}\  
\Bigg[
\langle\hat{X}_A(t) \hat{X}_A(t-\tau)\rangle \round{\hat{a}_\nu \rho_S \hat{a}_{\nu^\prime} - \ann{a}_{\nu^\prime} \ann{a}_\nu \rho_S}  +\langle\hat{X}_A(t) \hat{X}_B(t-\tau)\rangle \round{\hat{a}_\nu \rho_S \hat{b}_{\nu^\prime} - \ann{a}_{\nu^\prime} \ann{b}_\nu \rho_S} + \nonumber\\
 & +\langle\hat{X}_B(t) \hat{X}_A(t-\tau)\rangle \round{\hat{b}_\nu \rho \hat{a}_{\nu^\prime} - \ann{b}_{\nu^\prime} \ann{a}_\nu \rho} + 
 \langle\hat{X}_B(t) \hat{X}_B(t-\tau)\rangle \round{\hat{b}_\nu \rho \hat{b}_{\nu^\prime} - \ann{b}_{\nu^\prime} \ann{b}_\nu \rho} \Bigg] + \text{H.c.} \Bigg\}
\end{align}
We now rewrite this expression as
\begin{align}
\label{rhodot_Gammas}
\dot{\rho}(t) = 
- i \squared{\hat{H}_S^R,\rho(t)} +  \sum_{\nu,\nu^\prime} \Bigg\{e^{i \round{\nu+\nu^\prime}\omega_r t}\ \Bigg[
&\gamma_{AA}(\nu) \round{\hat{a}_\nu \rho \hat{a}_{\nu^\prime} - \ann{a}_{\nu^\prime} \ann{a}_\nu \rho} + 
 \gamma_{AB}(\nu) \round{\hat{a}_\nu \rho \hat{b}_{\nu^\prime} - \ann{a}_{\nu^\prime} \ann{b}_\nu \rho} +\nonumber\\
& +\gamma_{BA}(\nu) \round{\hat{b}_\nu \rho \hat{a}_{\nu^\prime} - \ann{b}_{\nu^\prime} \ann{a}_\nu \rho} + 
 \gamma_{BB}(\nu) \round{\hat{b}_\nu \rho \hat{b}_{\nu^\prime} - \ann{b}_{\nu^\prime} \ann{b}_\nu \rho} \Bigg] + \text{H.c.}\Bigg\}~,
\end{align}
where we defined 
\begin{align}
\label{Gammas_defs}
\gamma_{MN}(\nu) &= \int_0^\infty  d\tau e^{-i\omega_r\nu \tau}\langle \hat{X}_M(t) \hat{X}_N(t-\tau)\rangle \\
\end{align}
with $M, N$ running over $A$ and $B$, and the angular brackets indicating the average over the environment.

\subsection*{Two-mode squeezed environments}
\noindent We now include correlations in the remote environments $A$ and $B$, taking their modes to be initialized in two-mode squeezed states, so that
\begin{align}
\label{EqBathExpectations}
&\langle \crea{A}_\omega \ann{A}_{\omega^\prime}\rangle = \langle \crea{B}_\omega \ann{B}_{\omega^\prime}\rangle = \delta(\omega-\omega^\prime) n_r(\omega),\qquad
\langle \crea{A}_\omega \crea{A}_{\omega^\prime}\rangle = \langle \crea{B}_\omega \crea{B}_{\omega^\prime}\rangle = 0\\ 
&\langle \crea{A}_\omega \crea{B}_{\omega^\prime}\rangle  = \delta(\omega-\omega^\prime) m_r(\omega), \qquad
\langle \crea{A}_\omega \ann{B}_{\omega^\prime}\rangle  = 0,
\end{align}
while all single-quadrature expected values vanish for vacuum squeezed states.
Notice that locally the modes have the statistics of incoherent light with effective thermal excitations given by $N_\omega$, while the correlations are given by $M_\omega$. We find
\begin{equation}
\gamma_{AA}(\nu) = \int_0^\infty d\tau\  e^{-i\nu \omega_r \tau}
\int_0^\infty d\omega \frac{g^2(\omega)}{2\pi} \squared{e^{i\omega \tau} n_r(\omega) + e^{-i\omega \tau}\round{1+n_r(\omega)}  }.
\end{equation}
We now use $\int_0^\infty d\tau\ e^{-i x \tau}/\pi = \delta(x) - (i/\pi) \PV\round{1/x}$. We neglect here the principal value, which results in a Lamb shift that can be reabsorbed in the definition of the resonator's frequency. This simplification can be proven to be exact when the environmental cutoff frequency is sufficiently large~\cite{correa_potential_2025}. We find
\begin{equation}
\label{Gamma_AA}
\gamma_{AA}(\nu) = \gamma\squared{ n_r({\omega_r}) \Theta(\nu) + \round{1 + n_r({\omega_r})}\Theta(-\nu)},\ \quad 
\Theta(\nu) = \left\{
\begin{matrix}
1\ \text{if}\ \nu>0\\
0\ \text{if}\ \nu<0
\end{matrix}
\right. ,
\end{equation}
where we defined the effective dissipation rate $2 \gamma = g(\omega_r)^2$. 
Similarly, from Eq.~\eqref{Gammas_defs} we have
\begin{equation}
\gamma_{AB}(\nu)= \int d\omega \Biggl\{
\squared{ \int_0^\infty d\tau \frac{1}{\pi} e^{-i(\omega_r \nu + \omega)\tau}}  e^{2i\omega t}\frac{g(\omega)^2}{2} m_r(\omega)^*
\ +\
\squared{ \int_0^\infty d\tau \frac{1}{\pi} e^{-i(\omega_r \nu - \omega)\tau}}  e^{-2i\omega t}\frac{g(\omega)^2}{2} m_r(\omega)
\Biggr\}.
\end{equation}
Following the same assumptions used for Eq.~\eqref{Gamma_AA} we land on
\begin{equation}
\label{Gamma_AB}
\gamma_{AB}(\nu) = \gamma\squared{e^{-2i \omega_r t} m_r({\omega_r}) \Theta(\nu) + e^{2i \omega_r t} m_r({\omega_r})^*\Theta(-\nu)}.
\end{equation}
By direct inspection we have that $\gamma_{AA}(\nu) = \gamma_{BB}(\nu)$ and $\gamma_{AB}(\nu) = \gamma_{BA}(\nu)$. 
We can now use these results in Eq.~\eqref{rhodot_Gammas} and eliminate by rotating-wave approximation 
all the terms that are fast-oscillating at frequencies of the order of $\omega_r$. Notice that this is where we make use of the energy scale $\Delta,\lambda \ll \omega_r,\omega_d$, as we assume that the dominant frequencies of the systems are given by $\omega_r$ and the contribution given by the Hamiltonian $\hat{H}_S$ is too small to make the neglected terms relevant.\\
The full master equation can be decomposed as
\begin{align}
    \dot{\rho}(t) &= 
- i \squared{\hat{H}_S^R,\rho(t)} -  D^{(A)}[\rho(t)]  + D^{(B)}[\rho(t)]  + D^{(AB)}[\rho(t)]~.
\end{align}
The dissipator relative to the local (effective) thermal dissipative bath is given by,
\begin{align}
D^{(A)}[\rho(t)] 
&= \sum_{\nu,\nu'} \delta_{\nu,-\nu'} \,
\gamma
\Big(
n_r({\omega_r})\,\theta(\nu) + (1+n_r({\omega_r}))\theta(-\nu)
\Big)
\left(
\ann{a}_\nu \rho \ann{a}_{\nu'} - \ann{a}_{\nu'} \ann{a}_\nu \rho + \text{H.c.}
\right) \\
&= 
\gamma (1+n_r({\omega_r}))
\left(
2 \ann{a} \rho \crea{a} - \curly{\crea{a}\ann{a},\rho}
\right)
+ 
\gamma n_r({\omega_r})
\left(
2\crea{a} \rho \ann{a} - \curly{\ann{a} \crea{a},\rho}
\right)~,
\end{align}
and similarly for $D^{(B)}[\rho(t)]$. The genuinely nonlocal part of the master equation is given by the term
\begin{align}
D^{(AB)}[\rho(t)]
&=
\sum_{\nu,\nu'}
e^{i(\nu+\nu')\omega_r t}
\gamma
\Big[
e^{-2i\omega_r t} m_r({\omega_r)}\theta(\nu)
+
e^{2i\omega_r t} m_r({\omega_r})^* \theta(-\nu)
\Big]
\Big[
\hat{a}_\nu \rho \hat{b}_{\nu'} - \hat{a}_{\nu'} \hat{b}_\nu \rho
\Big]
\\
&\qquad + (\hat{a} \leftrightarrow \hat{b}) + \text{H.c.}
\end{align}
The time dependence can be rearranged as
\begin{align}
D^{(AB)}[\rho(t)]
&=
\sum_{\nu,\nu'}
e^{i\omega_r t(\nu+\nu'-2)}
\gamma
m_r({\omega_r})\theta(\nu)
\Big(
\hat{a}_\nu \rho \hat{b}_{\nu'} - \hat{a}_{\nu'} \hat{b}_\nu \rho + \text{H.c.}
\Big)
\\
&\quad +
\sum_{\nu,\nu'}
e^{i\omega_r t(\nu+\nu'+2)}
\gamma
m_r({\omega_r})^*\theta(-\nu)
\Big(
\hat{a}_\nu \rho \hat{b}_{\nu'} - \hat{a}_{\nu'} \hat{b}_\nu \rho + \text{H.c.}
\Big)
\\
&\qquad + (\hat{a} \leftrightarrow \hat{b})~.
\end{align}
After rotating-wave approximation, we select only the resonant terms identified by the relations
\begin{align}
e^{i\omega_r t(\nu+\nu'-2)}
&\rightarrow \nu+\nu'-2=0
\quad \Rightarrow \quad
\nu=\nu'=+~,
\\
e^{i\omega_r t(\nu+\nu'+2)}
&\rightarrow \nu+\nu'+2=0
\quad \Rightarrow \quad
\nu=\nu'=-~.
\end{align}
Rearranging the terms in compact Lindblad form, we finally find
\begin{align}
D^{(AB)}[\rho(t)]
=
2\gamma m_r({\omega_r})
\Big(
\crea{a} \rho \crea{b}
+
\crea{b} \rho \crea{a}
-
\{\crea{a}\crea{b},\rho\}
\Big) +
2\gamma m_r^*({\omega_r})
\Big(\hat{a} \rho \hat{b} + \hat{b} \rho \hat{a} -
\{\hat{a}\hat{b},\rho\}
\Big)~.
\end{align}

\label{app:HP_spectrum}

\section{Gaussian Wigner functions}

\noindent In this appendix, we derive the Wigner functions of the system in the Gaussian approximation. Introducing the Wigner-Weyl phase-space representation of the density operator $\rho(t)$, the Wigner function is defined as
\begin{equation}
    W(x_a,p_a,x_a,p_a;t):=\frac{1}{4\pi^2}\int_{-\infty}^{\infty}dy_a\int_{-\infty}^{\infty} dy_b\;\; e^{-i (p_a y_a- p_b y_b)}\; \left\langle x_a+\frac{y_a}{2},x_b+\frac{y_b}{2}\,\Big|\,\rho(t)\,\Big|\,x_a-\frac{y_a}{2},x_b+\frac{y_b}{2}\right\rangle~,
\end{equation}
in terms of which the master equation \eqref{Lindblad_ab} reduces to the Fokker--Planck equation 
\begin{align}\label{e. Fokker-Planck ab}
    \partial_t W=&\sum_{i\in\{a,b\}}\left[-\partial_{x_i}(F_{x_i}W)-\partial_{p_i}(F_{p_i}W)+\gamma\left(n_r+\frac{1}{2}\right)\left(\partial_{x_i}^2+\partial_{p_i}^2\right)W\right]\nonumber\\
    &\;\;-2\gamma\left[\mathrm{Re}\,m_r\,(\partial_{x_a}\partial_{x_b}-\partial_{p_a}\partial_{p_b})+\mathrm{Im}\,m_r\,(\partial_{x_a}\partial_{p_b}+\partial_{x_b}\partial_{p_a})\right]W + QN(W)~,
\end{align}
where the drift coefficients read
\begin{equation}
\begin{dcases}
     F_{x_i}= - \big(\gamma+\mathrm{Im}\,\lambda\,\big) \,x_i+\big[\omega-\mathrm{Re}\,\lambda\,+\epsilon\, (x_i^2+p_i^2-2)\big]\,p_i~,\\
     F_{p_i}=-\big[\omega+\mathrm{Re}\,\lambda\,+\epsilon\, (x_i^2+p_i^2-2)\big]\,x_i - \big(\gamma-\mathrm{Im}\,\lambda\big)\, p_i~.
\end{dcases}
\end{equation}
and $QN(W)$ represents a "quantum-noise" term arising from the non-commutativity of the ladder operators, e.g., $[\crea{a},\ann{a}]=[\crea{b},\ann{b}]=1$, and reads
\begin{equation}
    QN(W)=-\frac{\epsilon}{4}\sum_{i\in\{a,b\}}\left(x_i\partial_{p_i}-p_i\partial_{x_i}\right)\left(\partial_{x_i}^2+\partial_{p_i}^2\right)W~.
\end{equation} 
In the main text, we considered the case $\lambda\in\mathbb{R}^+$ and $m_r=-\sqrt{n_r(n_r+1)}\in \mathbb{R}^-$.

In the presence of only quadratic terms, $\epsilon=0$, the steady state of Eq.~\eqref{e. Fokker-Planck ab} is a Gaussian distribution,
\begin{equation}\label{e. Gaussian W}
    W_{ss}(x_a,p_a,x_b,p_b)=\frac{1}{(2\pi)^2\sqrt{\det(K)}}\,e^{-\frac{1}{2}\sum_{ij}{(K^{-1})}_{ij}\,\big(z_i-\langle \hat{z}_i\rangle\big)\big(z_j-\langle\hat{z}_j\rangle\big)}\quad \mbox{with}\quad z_i\in\{x_a,p_a,x_b,p_b\}~.
\end{equation}
The covariance matrix $K$ identifies the quadratures of the operators $\hat{z}_i$ via $K_{ij}=\langle\{\hat{z}_i,\hat{z}_j\}\rangle/2-\langle \hat{z}_i\rangle \langle \hat{z}_j \rangle$ and can be written as
\begin{equation}
    K=\begin{pmatrix}
        K_{a} & K_{ab}\\
        K_{ab}^T & K_{b}
    \end{pmatrix}~,
\end{equation}
where, because of the symmetry \(a\leftrightarrow b\), the diagonal blocks are identical, and the off-diagonal one satisfies \(K_{ab}=K_{ab}^{T}\). Moreover, parity symmetry implies $\langle \hat{z}_i\rangle=0$. Inserting the expression \eqref{e. Gaussian W} in Eq.~\eqref{e. Fokker-Planck ab} for $\epsilon=0$, we obtain the Lyapunov equation
\begin{equation}\label{e. Lyap K}
    \partial_tK=F\,K+K\,F^T+2\gamma G
\end{equation}
with
\begin{equation}\label{e. F,G general}
    \begin{dcases}
        F=\mathbb{I}_2\otimes\left(i\omega\sigma_y-|\lambda|\,R_{\arg{\lambda}}\right)-\gamma\,\mathbb{I}_4\\
         G=\left(n_r+\frac{1}{2}\right)\mathbb{I}_4+|m_r|\,\sigma_x\otimes R_{-\pi/2-\arg{m_r}}
    \end{dcases}
\end{equation}
and where $\sigma_{x,y}$ are Pauli matrices and $R_\nu$ is the rotation matrix
\begin{equation}
    R_\nu=\begin{pmatrix}
        \sin\nu & \cos\nu\\
        \cos\nu&  -\sin\nu
    \end{pmatrix}~.
\end{equation}
The steady-state solution of Eq.~\eqref{e. Lyap K} reads
\begin{equation}\label{e. Ka}
    K_a=K_b=\frac{n_r+1/2}{\omega^2+\gamma^2-|\lambda|^2}\begin{pmatrix}
    \omega^2+\gamma^2-\omega\,\mathrm{Re}\,\lambda-\gamma\,\mathrm{Im}\,\lambda& \omega\,\mathrm{Im}\,\lambda-\gamma\,\mathrm{Re}\,\lambda\\
   \omega\,\mathrm{Im}\,\lambda-\gamma\,\mathrm{Re}\,\lambda & \omega^2+\gamma^2+\omega\,\mathrm{Re}\,\lambda+\gamma\,\mathrm{Im}\,\lambda
    \end{pmatrix}~,
\end{equation}
and $K_{ab}=K_{ab}^{R}+K_{ab}^{I}$ with
\begin{align}
    &K_{ab}^R=\frac{\mathrm{Re}\,m_r}{\omega^2+\gamma^2-|\lambda|^2}\begin{pmatrix}
   \gamma\left(\mathrm{Im}\,\lambda-\gamma\right)+\mathrm{Re}\,\lambda\left(\mathrm{Re}\,\lambda-\omega\right)& \omega\,\gamma-\mathrm{Re}\,\lambda\,\,\mathrm{Im}\,\lambda\\
  \omega\,\gamma-\mathrm{Re}\,\lambda\,\,\mathrm{Im}\,\lambda & \gamma\left(\mathrm{Im}\,\lambda+\gamma\right)-\mathrm{Re}\,\lambda\left(\mathrm{Re}\,\lambda+\omega\right)
    \end{pmatrix}~, \label{e. KabR}\\
  \nonumber \\
    &K_{ab}^I=\frac{\mathrm{Im}\,m_r}{\omega^2+\gamma^2-|\lambda|^2}\begin{pmatrix}
   -\left(\mathrm{Im}\,\lambda-\gamma\right)\left(\mathrm{Re}\,\lambda-\omega\right)& (\mathrm{Im}\,\lambda)^2-\gamma^2\\
  (\mathrm{Im}\,\lambda)^2-\gamma^2 & \left(\mathrm{Im}\,\lambda+\gamma\right)\left(\mathrm{Re}\,\lambda+\omega\right)
    \end{pmatrix}~.\label{e. KabI}
\end{align}
The eigenvalues of the matrix $F$ in Eqs.~\eqref{e. F,G general} reads $\mu_{\pm}=-\gamma\pm\sqrt{|\lambda|^2-\omega^2}$ and have positive real part only for $|\lambda|<\sqrt{\omega^2+\gamma^2}$, meaning that the solution \eqref{e. Ka}-\eqref{e. KabR}-\eqref{e. KabI} is stable only if the driving strength stays below the critical threshold $\lambda_c=\sqrt{\omega^2+\gamma^2}$. Taking $\lambda \in \mathbb{R}$ and $m_r=-\sqrt{n_r(n_r+1)}$, we obtain the steady-state covariance matrix $\Sigma$ solving Eq.~\eqref{e. cov matrix equation} in Sec.~\ref{s. Gaussian regime}, \textit{i.e.} 
\begin{equation}\label{e. Sigma}
   \Sigma_a=\Sigma_b=\frac{1}{\lambda_c^2-\lambda^2}\left(n_r+\frac{1}{2}\right)
   \begin{pmatrix}
        \lambda_c^2-\omega\lambda & -\gamma\lambda\\
        -\gamma\lambda & \lambda_c^2+\omega\lambda
    \end{pmatrix}~,\quad\Sigma_{ab}=\frac{\sqrt{n_r(n_r+1)}}{\lambda_c^2-\lambda^2}\begin{pmatrix}
   \gamma^2-\lambda\left(\lambda-\omega\right)& -\omega\,\gamma\\
  -\omega\,\gamma & -\gamma^2+\lambda\left(\lambda+\omega\right)
    \end{pmatrix}~.
\end{equation}
The matrix $\Sigma_a=\Sigma_b$ describes the marginal distributions of the Wigner function in the local planes $x_ap_a$ and $x_bp_b$. It defines a covariance ellipse with an eccentricity 
$\sqrt{2\lambda/(\lambda+\lambda_c)}$, rotated by an angle $\xi$ with respect to the $x$ axis, where
\begin{equation}\label{e. squeezing direction}
    \tan\xi=\frac{\gamma}{\omega-\sqrt{\gamma^2+\omega^2}}.
\end{equation}
As $\lambda$ approaches $\lambda_c$, the ellipse becomes increasingly squeezed along the direction $\xi$ until it degenerates into an infinite straight band. In the limit $\omega\gg\gamma$, this local squeezing direction approaches $\pi/2$, implying that the degeneration occurs along the $p$ axis. Moving to the $p_ap_b$ plane, according to Eq.~\eqref{e. Sigma}, the covariance matrix describing the marginal distribution reads

\begin{equation}\label{e. Sigma}
   \Sigma_{p_ap_b}=\frac{1}{\lambda_c^2-\lambda^2}
   \begin{pmatrix}
\left(n_r+1/2\right)\left(\lambda_c^2+\omega\lambda\right) & \sqrt{n_r(n_r+1)}\left(-\gamma^2+\lambda^2+\lambda\omega\right)\\
\sqrt{n_r(n_r+1)}\left(-\gamma^2+\lambda^2+\lambda\omega\right) & \left(n_r+1/2\right)\left(\lambda_c^2+\omega\lambda\right)
\end{pmatrix}~.
\end{equation}
It defines a covariance ellipse rotated by $\pi/4$, reflecting the symmetry $a\leftrightarrow b$, whose eccentricity, for $\omega \gg \gamma$, increases with $\lambda$ and approaches $1-e^{-4r}$ as $\lambda \to \lambda_c$. Consequently, for strong squeezing, the covariance ellipse in the $p_a p_b$ plane becomes increasingly degenerate.\\

\noindent As described in Sec.~\ref{s. Gaussian regime}, for $|\lambda|>\lambda_c$ the Wigner function has four distinct semiclassical equilibrium points. According to Eqs.~\eqref{e. H2 above}-\eqref{e. Omega, Lambda}, Gaussian fluctuations around a single point are described by the Fokker-Planck equation \eqref{e. Fokker-Planck ab} with the replacements $\omega\rightarrow \Omega$ and $\lambda\rightarrow\Lambda\notin\mathbb{R}$. Accordingly, the Wigner function is given by Eq.~\eqref{e. Gaussian W} with $\langle \hat{z}_i \rangle$ given by Eq.~\eqref{e. semiclassical solutions}. In this way, the covariance matrix of each Gaussian solution can be obtained from Eqs.~\eqref{e. Ka}--\eqref{e. KabR}--\eqref{e. KabI}. Each solution turns out to be stable for $\lambda > \lambda_c$, as the eigenvalues of the new matrix $F$ are $-\gamma \pm \sqrt{5\gamma^2 - 4\lambda^2 + 4\omega \sqrt{\lambda^2 - \gamma^2}}$ and have strictly negative real parts in this regime. Considering the local planes $x_ap_a$ or $x_bp_b$ and the plane $p_a p_b$, an analysis based on Eq.~\eqref{e. H2 above} can be performed, analogous to that carried out below threshold, to obtain the eccentricity of the covariance ellipses describing fluctuations around each semiclassical equilibrium point. Consistently, these ellipses become increasingly degenerate as $\lambda \to \lambda_c$ from above.

\section{Collective modes}

\noindent Introducing the collective modes $\ann{c}=(\ann{a}+\ann{b})/\sqrt{2}$, $\ann{d}=(\ann{a}-\ann{b})/\sqrt{2}$, the environmental correlations are decoupled and the master equation \eqref{Lindblad_ab} becomes

\begin{equation}\label{e. master cd}
    \partial_t\rho=-i\,[H_{int},\rho]+\mathcal{L}_c[\rho]+\mathcal{L}_d[\rho]~,
\end{equation}
where $\mathcal{L}_c[\rho]=-i [H_c,\rho]+ \mathcal{D}_c^{th}[\rho]+\mathcal{D}_c^{sq}[\rho]$ with
\begin{align}
    \begin{dcases}
        H_c=\omega c^\dagger c+\frac{1}{2}\left(\lambda \,c^2 + \lambda^* c^{\dagger 2}\right)~,\\
        \mathcal{D}_c^{th}[\rho]=\gamma (1+n_r) \left(2c\rho c^\dagger- \{c^\dagger c,\rho\}\right)+\gamma n_r \left(2c^\dagger\rho c- \{c c^\dagger,\rho\}\right)~,\\
        \mathcal{D}_c^{sq}[\rho]=\gamma m_r \left(2c^\dagger\rho c^\dagger- \{c^{\dagger 2},\rho\}\right)+\gamma m_r^* \left(2c\rho c- \{c^2,\rho\}\right)~,
        \end{dcases}
\end{align}
$\mathcal{L}_d[\rho]=\mathcal{L}_c|_{m_r\rightarrow -m_r}[\rho]$ and
\begin{equation}\label{e. Hint}
    H_{int}=\frac{\epsilon}{2}\left(c^{\dagger 2}+d^{\dagger 2}\right)\left(c^2+d^2\right)+\epsilon \,c^\dagger c\, d^\dagger d~.
\end{equation}
Notice that the new master equation \eqref{e. master cd} does not contain correlated dissipative terms between the collective modes $\hat{c}$ and $\hat{d}$, but it contains the Hamiltonian non-linear interaction \eqref{e. Hint}. In the Wigner-Weyl representation, the master equation \eqref{e. master cd} is mapped to the Fokker-Planck equation
\begin{align}
\partial_t W=&\sum_{j\in\{c,d\}}\left[-\partial_{x_j}(F_{x_j}W)-\partial_{p_j}(F_{p_j}W)+\frac{1}{2}\left(Q_{x_j}\partial_{x_j}^2+Q_{p_j}\partial_{p_j}^2+2\,Q_{{x_jp_j}}\,\partial_{x_j}\partial_{p_j}\right)W\right]+ QN(W)~,
    \label{e. Fokker cd}
\end{align}
where the drift coefficients read
\begin{equation}\label{e. drift coeff}
\begin{dcases}
    F_{x_j}= - \big(\gamma+\mathrm{Im}\,\lambda\,\big) \,x_j+\big(\omega-\mathrm{Re}\,\lambda\big)\,p_j\,+\frac{\epsilon}{2}\left[p_j^3+2\,x_j\,x_{\overline{j}} \,\,p_{\overline{j}}+p_j\,(x_j^2+x_{\overline{j}}^2+3\,p_{\overline{j}}^2-4)\right]~,\\
     F_{p_j}=-\big(\omega+\mathrm{Re}\,\lambda\big)\,x_j- \big(\gamma-\mathrm{Im}\,\lambda\big)\, p_j~-\frac{\epsilon}{2}\left[x_j^3+2\,p_j\, p_{\overline{j}}\,\,x_{\overline{j}}+x_j\,(p_j^2+p_{\overline{j}}^2+3\,x_{\overline{j}}^2-4)\right]~,
    \end{dcases}
\end{equation}
and, for the sake of a lighter notation, we used a bar $\bar j $ to denote index exchange $c\leftrightarrow d$, as in the main text. The influence of the environmental correlations, present for $m_r\neq 0$, is here expressed as anisotropic diffusion, giving rise to distinct diffusion rates along the $x$ and $p$ directions, \textit{i.e.}
\begin{equation}\label{e. QxQp}
\begin{dcases}
    Q_{x_c}=Q_{p_d}=\gamma\,(1+2n_r-2\,\mathrm{Re}\,m_r)
    ~,\\
    Q_{x_d}=Q_{p_c}=\gamma\,(1+2n_r+2\,\mathrm{Re}\,m_r)
    ~,\\
    Q_{x_cp_c}=-Q_{x_dp_d}=2\,\gamma\,\mathrm{Im}\,m_r~.
\end{dcases}
\end{equation}
Note that, clearly, Eq.~\eqref{e. Fokker cd} can be obtained from Eq.~\eqref{e. Fokker-Planck ab} via the change of variables $x_c=(x_a+x_b)/\sqrt{2}$, $x_d=(x_a-x_b)/\sqrt{2}$ and similar for $p$.

\section{Adiabatic elimination}
\noindent The eigenvalues $\mu_\pm$ of the Gaussian dynamical matrix $F$ of Eq.~\eqref{e. F,G general} show a separation of timescales when $|\lambda|\to\lambda_c$. In particular, as underlined at the beginning of Sec.~\ref{s. slaving}, when the critical point is approached $\mu_+$ vanishes, resulting in a critical slowing down, while $\mu_-\to-2\gamma$ stays finite. The variables $u$ and $v$ diagonalizing $F$ read
\begin{equation}\label{e. u,v def general}
\begin{dcases}
u=-x \cos\chi + p \sin\chi\\
v = x \cos\left(\chi-\arg{\lambda}\right) + p \sin\left(\chi-\arg{\lambda}\right)
\end{dcases}~,
\qquad
\tan\left(\chi-\frac{\arg{\lambda}}{2}\right) = \sqrt{\frac{|\lambda| - \omega}{|\lambda| + \omega}}~.
\end{equation}
The variable $v$ is associated with $\mu_-$ and can then be regarded as a \textit{fast} variable, while the variable $u$ is critically \textit{slow} as it is associated with the vanishing eigenvalue $\mu_+$. Note that, consistently, for $\lambda\in\mathbb{R}$ Eq.~\eqref{e. u,v def general} matches Eq.~\eqref{e. u,v def}.
Neglecting the quantum noise~\cite{dykman2012fluctuating}, the Fokker-Planck equation \eqref{e. Fokker cd} can be written as a continuity equation $\partial_t W+\sum_j\partial_{u_j} J_{u_j}=0$ where, in terms of $u$ and $v$, the probability currents read

\begin{align}
    \begin{dcases}
        J_{v_j}=F_{v_j}W-\frac{1}{2}\left(Q_{v_j}\,\partial_{v_j}+2\,Q_{u_jv_j}\,\partial_{u_j}\right)W\\
        J_{u_j}=F_{u_j}W-\frac{1}{2}Q_{u_j}\,\partial_{v_j}W
    \end{dcases}
\end{align}
with 
\begin{equation}\label{e. FQ uv}
    \begin{dcases}
        F_{u}=-F_{x} \cos\chi +F_{p} \sin\chi\\
        F_{v}=F_x \cos\left(\chi-\arg{\lambda}\right) + F_p \sin\left(\chi-\arg{\lambda}\right)
    \end{dcases}~,\quad\begin{dcases}
      Q_{u}= Q_x \cos^2\!\chi+Q_p\sin^2\!\chi - Q_{xp}\sin 2\chi~,\\
      Q_{v}= Q_x \cos^2\!\left(\chi-\arg{\lambda}\right)+Q_p\sin^2\!\left(\chi-\arg{\lambda}\right) + Q_{xp}\sin[2\!\left(\chi-\arg{\lambda}\right)]~,\\
    Q_{uv}= -Q_x \cos\!\left(\chi-\arg{\lambda}\right)\cos\!\chi+Q_p\sin\!\left(\chi-\arg{\lambda}\right)\sin\!\chi+Q_{xp}\sin(\arg{\lambda})~,
    \end{dcases}
\end{equation}
\vspace{-1mm}\\
and the above relations apply to both modes $c$ and $d$; the mode subscripts $j$ in Eqs.~\eqref{e. FQ uv} have been omitted to keep the notation lighter. \\
As a consequence of the critical slowing down, in the limit $|\lambda| \to \lambda_c$, the variables $v_j$ relax to their steady values much faster than $u_j$. This implies that the long-time behavior of $W$ is governed entirely by $u_j$. In fact, close to the steady state, the variables $v_j$ have already relaxed to their stationary value 
and no longer exhibit independent dynamics; instead, they are slaved to follow the slow evolution of $u_j$. This timescale separation enables the use of \textit{adiabatic elimination}, which allows for a reduction in the number of effective variables. A general discussion of this framework can be found in Ref.~\cite{hutt_synergetics_2020}; in the following, we briefly summarize its application to the system of interest.

\begin{itemize}

\item After a fast initial transient, the distribution of the variables $v_j$ relaxes to a local equilibrium form $h_j(v_j|u_c,u_d)$, depending on 
$u_c$ and $u_d$. This distribution is stationary on the fast timescale and adjusts adiabatically to the slowly evolving $u_j$. Consequently, the $u_j$-dependence of $h_j$ is weak compared to its variation with respect to $v_j$, and the fast variables remain constrained to follow the dynamics dictated by the slower ones. The function $h_j$ identifies the conditional probability for $v_j$ given $u_c$ and $u_d$ and corresponds to a time-independent one-dimensional Gibbs distribution. This is derived by imposing vanishing probability current along the $v_j$-direction, while treating $u_j$ as a static external parameter, namely $J_{v_j}=0$ with $\partial_{u_j} h_j \ll \partial_{v_j} h_j$, yielding 

\begin{equation}\label{e. hj}
  h_j(v_j|u_c,u_d) \propto \exp \left[ \frac{2}{Q_{v_j}} \int \!\!dv_j \, F_{v_j} \right].
\end{equation}

\item The time evolution of the full Wigner function $W$ is thus entirely determined by the reduced marginal distribution $K(u_c,u_d,t)$, with the variables $v_j$ serving as rapidly fluctuating noise sources, whose mean effect is captured by averaging the drift terms $F_{u_j}$ over the local equilibrium distribution $h_j$. The distribution $K$ satisfies thus the reduced Fokker-Planck equation
\begin{equation}\label{e. reduced Fokker}
    \partial_t K+\sum_{j\in\{c,d\}}\partial_{u_j}\langle J_{u_j} \rangle=0
\end{equation}
with $\langle J_{u_j} \rangle$ the probability current along the slow direction averaged on the noise given by the fast $v_j$ dynamics reading
\begin{equation}\label{e. K equation sup}
    \langle J_{u_j} \rangle=\langle F_{u_j}\rangle K-\frac{1}{2}Q_{u_j}\partial_{u_j}K\quad\mbox{with}\quad
    \langle F_{u_j} \rangle=\int\! dv_j\,F_{u_j}\,h_j~.
\end{equation}

\item Within this framework, we write the Wigner function as in Eq.~\eqref{e. W slaving}:
\begin{equation}\label{e. W slaving sup}
    W=h_c(v_c|u_c,u_d) \;h_d(v_d|u_c,u_d)\; K(u_c, u_d,t)~,
\end{equation}
with the normalizations conditions $\int\!dv_{j}\, h_j(v_j|u_c,u_d)=\int\! du_{c}du_{d}\,K(u_c, u_d,t)=1$, and use Eqs.~\eqref{e. hj} and \eqref{e. K equation sup} to find the steady state.
\end{itemize}
An explicit expression for the drift terms can be obtained via Eqs.~\eqref{e. FQ uv}, \eqref{e. drift coeff} and \eqref{e. u,v def}:
\begin{equation}\label{e. F sup}
    \begin{dcases}
        F_{v_j}=\mu_+ v_j+\frac{\epsilon\,|\lambda|\, u_j}{2\sqrt{|\lambda|^2-\omega^2}} \left[\frac{|\lambda|^2}{|\lambda|^2-\omega^2}\left(u_j^2+3u_{\overline{j}}^2\right)-4\right]+\epsilon\,o(v_c,v_d)\\
        F_{u_j}=\mu_-\,u_j-\frac{\epsilon\,\omega\, u_j}{2\sqrt{|\lambda|^2-\omega^2}} \left[\frac{|\lambda|^2}{|\lambda|^2-\omega^2}\left(u_j^2+3u_{\overline{j}}^2\right)-4\right]+\epsilon\,o(v_c,v_d)
    \end{dcases}
\end{equation}
where we reported only the first orders in $\epsilon$ and $v_c$, $v_d$. From Eq.~\eqref{e. F sup}, it is evident that, as $|\lambda| \to \lambda_c$, a separation of timescales emerges between $v_j$ and $u_j$. Indeed, according to $2\gamma\sim|\mu_+|\gg|\mu_-|\sim 0$, it is $2\gamma\sim\partial_{v_j} F_{v_j}\gg \partial_{u_j} F_{u_j}\sim\epsilon$. Because of the different timescales, the marginal distribution $K$ can be regarded as nearly constant relative to the conditional distributions $h_j$. The latter can be approximated as Gaussian functions sharply localized around their maxima $v_j^{s}$ satisfying $F_{v_j^s}=0$, namely
\begin{align}\label{e. conditional h sup}
h_j(v_j|u_{c},u_{d})= 
\sqrt{\frac{2\gamma}{\pi\, Q_{v_j}}} \;
e^{-\tfrac{2\gamma}{Q_{v_j}} \left[v_{j} - v_{j}^{ad}(u_{c}, u_{d})\right]^2}~,
\end{align}
where $v_j^{ad}$ denotes the "slaved" instantaneous equilibrium value of $v_j$ due to the slow variables dynamics adiabatic adjustment. 
For $|\lambda|\sim\lambda_c$ and $1\ll\omega/\gamma\ll\gamma/\epsilon$, Eq.~\eqref{e. F sup} becomes Eq.~\eqref{e.FuFv} and implies, at first order in $\epsilon$, 
\begin{equation} \label{e. vjs}
    v_j^{ad}(u_c,u_d) = \left[ \beta \left( u_j^2 + 3 u_{\bar{j}}^2 \right) - \zeta \right] \frac{u_j}{2\gamma}~,
\end{equation}
with $\beta,\zeta$ defined below Eq.~\eqref{e.FuFv}. In the limits $|\lambda|\sim\lambda_c$ and $\omega\gg\gamma$, Eq.~\eqref{e. u,v def general} gives $2\chi\sim\arg{\lambda}$, meaning that, according to Eqs.~\eqref{e. QxQp} and \eqref{e. FQ uv}, it is $Q_{v_j}=Q_{u_j}=-Q_{u_jv_j}:=Q_j$ with
\begin{equation}\label{e. QcQd}
    \begin{dcases}
        Q_c=\gamma\left[1+2n_r-2|m_r|\cos(\arg{\lambda}-\arg{m_r})\right]~,\\
        Q_d=\gamma\left[1+2n_r+2|m_r|\cos(\arg{\lambda}-\arg{m_r})\right]~.
    \end{dcases}
\end{equation}
Consistently, for $\arg{\lambda}=0$ and $\arg{m_r}=\pi$, and by inserting $n_r=\sinh^2{r}$ and $|m_r|=\sinh{r}\cosh{r}$, $Q_j$ reads as in Eq.~\eqref{e. QuQv}.\\
The averaged drift $\langle F_{u_j}\rangle$ can now be obtained via Eqs.~\eqref{e. K equation sup}, \eqref{e. F sup} and \eqref{e. conditional h sup}. However, at first order in $\epsilon$, it is $\langle F_{u_j}\rangle=F_{u_j}$ and the fast variables noise gives no contribution to the slow dynamics. Finally, for $|\lambda|\sim\lambda_c$ and $\omega\gg\gamma$, the reduced Fokker-Planck equation \eqref{e. reduced Fokker} describing the marginal distribution $K(u_c,u_d,t)$ can be written as in Eqs.~\eqref{e.Fokker_K}-\eqref{e.potential}.

\section{Correlation-induced adiabatic elimination}
\noindent Let us now focus on the steady state of the distribution $K$.  In the absence of bath correlations $(M=0)$, the diffusion coefficients are the same, \textit{i.e.} $Q_c=Q_d=Q$, and $K$ relaxes to the Gibbs state
\begin{equation}\label{e. Kssr0}
  K_{ss}(u_c,u_d) \propto \exp \left[-\frac{2}{Q}\,U(u_c,u_d) \right] \quad \mbox{with}\quad \langle F_{u_j}\rangle=-\,\partial_{u_j}U(u_c,u_d)~.
\end{equation}
However, non-vanishing bath correlations preclude an equilibrium
Gibbs state; indeed, $m_r\neq 0$ implies unequal diffusion coefficients.
Nevertheless, if one of the two coefficients is significantly larger than the other, the reduced Fokker–Planck equation \eqref{e. reduced Fokker} can again be solved via
adiabatic elimination. According to Eq.~\eqref{e. QcQd}, and being $|m_r|=\sinh{r}\cosh{r}$, this happens when the bath correlations are strong and in-phase or out-of- phase with the local drivings, \textit{i.e} if $\arg{\lambda}-\arg{m_r}= 0,\pi$ and $e^{4r}\gg 1$. In this case, the system exhibits an additional timescale separation: the variable with a higher diffusion rate moves easily between the potential wells, settling into its final state long before the other. It is noteworthy that this further timescale separation arises purely from stochastic reasons, as a consequence of the asymmetry in the diffusion coefficients determined by the strong bath correlations. In contrast, the timescale separation between the variables $u_j$ and $v_j$ has an intrinsic deterministic nature and would persist even in the absence of correlations. Accordingly, for $Q_c\gg Q_d$, we write the reduced distribution $K$ as in Eq.~\eqref{e. K decomposition}:
\begin{equation}\label{e. K decomposition sup}
    K(u_c,u_d,t)= k(u_c|u_d)\,y(u_d,t)
\end{equation}
with the normalization conditions $\int\!du_{c}\, k(u_c|u_d)=\int\! du_{d}\,y(u_d,t)=1$, and the distribution $k(u_c|u_d)$ defining the conditional probability of $u_c$ given $u_d$. This is the Gibbs distribution obtained by imposing vanishing probability current $\langle J_{u_c}\rangle=0$ in Eq.~\eqref{e. reduced Fokker}.
To first order in $\epsilon$, and for $|\lambda|\sim\lambda_c$ with $1\ll\omega/\gamma\ll\gamma/\epsilon$, 
the function $k$ reads as in Eq.~\eqref{e. conditional k Gibbs}, where, if $\lambda\notin\mathbb{R}$, $\lambda$ needs to be replaced by its modulus.
From the perspective of $u_d$, the variable $u_c$ behaves as a fluctuating noise source, resulting in an effective one-dimensional Fokker–Planck equation for $y(u_d,t)$:
\begin{equation}
    \partial_t y+\partial_{u_c}\langle\!\langle J_{u_d}\rangle\!\rangle=0
\end{equation}
where $\langle\!\langle J_{u_d}\rangle\!\rangle$ is the probability current $J_{u_d}$ averaged on the noise given both by $v_d$ and $u_c$, \textit{i.e.} 
\begin{equation}
   \langle\!\langle J_{u_d}\rangle\!\rangle=\langle\!\langle F_{u_d}\rangle\!\rangle y-\frac{1}{2} Q_{u_d}\partial_{u_d} y\quad\mbox{with}\quad \langle\!\langle F_{u_d}\rangle\!\rangle=\int\! d u_c\, \langle F_{u_d}\rangle\, k=\int\!\!\!\!\!\int\!du_c\, dv_d\,\, F_{u_d}\, k\, h_d~.
\end{equation}
To first order in $\epsilon$, one has $\langle F_{u_d}\rangle = F_{u_d}$. It then follows from Eq.~\eqref{e. average drift Fud} that the stationary condition $\langle\!\langle J_{u_d}\rangle\!\rangle = 0$ yields the steady-state distribution $y_{\mathrm{ss}}(u_d) \propto [\mathcal{N}(u_d)]^{Q_c/Q_d}$.
The factor $\mathcal{N}(u_d)$ can be obtained via the integral \eqref{e. integral N} which, considering Eq.~\eqref{e.potential}, reads
\begin{equation}\label{e. normalization factor 1}
   \mathcal{N}(u_d)=\exp{\left[\frac{1}{Q_c}\left(\alpha\,u_d^2-\frac{\beta}{2}u_d^4\right)\right]}\;\;\int_{-\infty}^{+\infty}du_c\,\exp{\left[-\left(\frac{3\,\beta\,u_d^2-\alpha}{Q_c}\right)\,u_c^2- \left(\frac{\beta}{2Q_c} \right)\,u_c^4\right]}
\end{equation}
with $\alpha$ and $\beta$ defined below Eq.~\eqref{e.FuFv}. Equation~\eqref{e. normalization factor 1} can be rewritten as
\begin{equation}\label{e. normalization factor 2}
    \mathcal{N}(u_d)=\exp{\left[\frac{1}{Q_c}\left(\alpha\,u_d^2-\frac{\beta}{2}u_d^4\right)\right]}\;\; \sqrt{\pi}\,\left(\frac{2Q_c}{\beta}\right)^{1/4} H_{-1/2}\left(\frac{3\beta u_d^2-\alpha}{\sqrt{2\beta Q_c}}\right)~,
\end{equation}
where $H_\nu(z)$ are Hermite functions with negative index, defined by 
\begin{equation}
    H_\nu(z)=\frac{1}{\Gamma(-\nu)}\,\int_{0}^{\infty}ds\,\,e^{-(s^2+2zs)}\,s^{-(\nu+1)}
\end{equation}
with $\mathrm{Re}(\nu)<0$ and $\Gamma(-\nu)$ Euler gamma function.
Finally, considering Eqs.~\eqref{e. K decomposition sup}, \eqref{e. conditional k Gibbs} and \eqref{e. normalization factor 2}, the steady state for $K$ reads
\begin{equation}\label{e. Kss ucud sup}
    K_{ss}(u_c,u_d)=\frac{1}{Z}\,\exp\left[-\frac{2}{Q_c} U(u_c,u_d)\right] \,J(u_d^2)~,
\end{equation}
with $Z$ normalization constant. We also defined the function
\begin{equation}
    J(u_d^2)=\exp\left\{\left(\frac{Q_c}{Q_d}-1\right)\left[\frac{1}{Q_c}\left(\alpha\,u_d^2-\frac{\beta}{2}u_d^4\right)+\log{H_{-1/2}\left(\frac{3\beta u_d^2-\alpha}{\sqrt{2\beta Q_c}}\right)}\,\right]\,\right\}~,
\end{equation}
describing a uni-modal symmetric curve peaked around $u_d=0$. It can be approximated by expanding the exponent at first order in $u_d^2$:
\begin{equation}\label{e. R approx}
    J(u_d^2)\sim\exp\left[-\,\sqrt{\frac{2\beta}{Q_c}}\left(\frac{Q_c}{Q_d}-1\right)\,f\!\left(\frac{\alpha}{\sqrt{2\beta Q_c}}\right)\,u_d^2\right]\quad\mbox{with}\quad f(w)=\frac{3\,H_{-3/2}(-w)}{2\,H_{-1/2}(-w)}-w~.
\end{equation}
The approximate expression for $K_{ss}(u_c,u_d)$ given in Eq.~\eqref{e. Kss ucud} follows by inserting the explicit forms of $Q_c,Q_d$ in Eq.~\eqref{e. R approx}, and by noting that, near the threshold, $\alpha \sim 0$ so that $f(w) \sim 1$.
As noted in Sec.~\ref{s. slaving}, strong bath correlations imply $Q_c\gg Q_d$, meaning that $K_{ss}$ is a Gibbs distribution induced by the potential $U$ and supplemented by a delta-like correction sharply localizing it in the $u_c$-direction.\\
Let us, however, underline that for $Q_c=Q_d$ the function $J(u_d^2)$ is identically equal to one, meaning that Eq.~\eqref{e. Kss ucud sup} (and the approximated Eq.~\eqref{e. Kss ucud}) identifies the correct steady state \eqref{e. Kssr0} also in the absence of bath correlations ($m_r=0$).  This may appear surprising, as we have emphasized that the adiabatic elimination can be applied only when the two diffusion coefficients are strongly asymmetric. Actually, although the adiabatic elimination breaks down for $Q_c\sim Q_d$, the conditions it suggests remain accurate. In fact, Eq.~\eqref{e. Kss ucud sup} was obtained by imposing $\langle\!\langle J_{u_d}\rangle\!\rangle = 0$ and $\langle J_{u_c}\rangle = 0$. For a confining potential, the first condition holds identically, as follows from integrating \eqref{e. reduced Fokker} over $u_c$. The second condition can provide a controlled approximation for two different reasons: \textit{(i)} if $Q_c\gg Q_d$, $u_c$ relaxes faster than $u_d$ and the associated current is negligible near the steady-state; \textit{(ii)} if $Q_c\sim Q_d$, the system reaches an approximate global equilibrium, namely a Gibbs state where all the fluxes are zero. Finally, Eq.~\eqref{e. Kss ucud sup} accurately captures the steady-state distribution $K_{ss}$ both for strong and small bath correlations, \textit{i.e.}, both for $e^{4r}\gg 1$ and $e^{4r}\ll 1$.

\section{The marginal distribution in the $p_ap_b$ plane}

\noindent We are now interested in exploiting the solutions \eqref{e. conditional h sup} and \eqref{e. Kss ucud sup} to derive the steady-state marginal distribution $W_{ss}(p_a,p_b)$ in the $p_ap_b$ plane via Eq.~\eqref{e. W slaving sup}. To this purpose,  we use Eq.~\eqref{e. u,v def general} to map the constraints $v_j = v_j^s$, imposed by the sharply localized Gaussian conditional distributions \eqref{e. conditional h sup}, onto constraints $x_j = x_j^s$ relating the variables $x_j$ to $p_j$. For $|\lambda|\sim\lambda_c$ and $\omega\gg\gamma$, Eq.~\eqref{e. u,v def general} gives $2\chi\sim\arg{\lambda}+\gamma/\omega$ so that, at first order in $\epsilon$, it is
\begin{equation}
    x_j^s= - p_j \tan{\chi}\left[1+\frac{\zeta}{\gamma}-\frac{4\beta \sin^2{\chi}}{\gamma}\left(p_j^2+3p_{\overline{j}}^2\right)\right]~.
\end{equation}
We can thus obtain the marginal distribution in the $p_cp_d$ plane as
\begin{equation}
    W_{ss}^{\,cd}(p_c,p_d)=K_{ss}(u_c,u_d)\,\Big|_{\,u_j\,=\,-x_j^s\cos\chi+p_j \sin\chi\,=\,2\,p_j\sin\chi\,+\,o(\epsilon)}
\end{equation}
which, rewriting the approximate form of Eq.~\eqref{e. Kss ucud sup} as
\begin{equation}
    K_{ss}(u_c,u_d)=\frac{1}{Z}\,\exp\left[-\frac{2}{Q_c} \,U_{\mathrm{eff}}^{\,cd}(u_c,u_d)\right]\quad\mbox{with}\quad U_{\mathrm{eff}}^{\,cd}(u_c,u_d)=U(u_c,u_d)+\xi\, u_d^2~,\;\;\xi=\sqrt{\frac{\beta Q_c}{2}}\left(\frac{Q_c}{Q_d}-1\right)
\end{equation}
reads
\begin{equation}
    W_{ss}^{\,cd}(p_c,p_d)=\frac{1}{Z}\,\exp\left[-\frac{2}{Q_c} \,U_{\mathrm{eff}}^{\,cd}\,\Big(2p_c\sin\chi,2p_d\sin\chi\Big)\right]~.
\end{equation}
Now, according to the definition of $\ann{c}$ and $\ann{d}$, it is
\begin{equation}\label{e. Wsspapb sup}
    W_{ss}(p_a,p_b)=W_{ss}^{\,cd}\left(\frac{p_a+p_b}{\sqrt{2}},\frac{p_a-p_b}{\sqrt{2}}\right)=\frac{1}{Z}\,\exp\left[-\frac{1}{D} \,U_{\mathrm{eff}}\Big(2p_a\sin\chi,2p_b\sin\chi\Big)\right]~,
\end{equation}
with $D=Q_c/2$ and $U_{\mathrm{eff}}(p_a,p_b)$ defined as in Eq.~\eqref{e. Ueff}. Note that in the main text we took $\arg\lambda=0$, implying $\chi\sim 2\gamma/\omega\sim 0$ and $\sin\chi\sim\chi$. Moreover, for also $\arg M=\pi$, we used expression \eqref{e. QuQv} to write $\xi$ as below Eq.~\eqref{e. Ueff}. For the sake of completeness, we write below the general expressions for $\xi$ and $D$ in terms of $n_r$ and $m_r$:
\begin{equation}
\begin{dcases}
    D=\frac{\gamma}{2}\,\big[1+2n_r-2m_r\cos({\arg\lambda-\arg m_r})\big]\\
    \xi=-\,\sqrt{8\beta\gamma}\;\,m_r\,\cos({\arg\lambda-\arg m_r})\,\,\frac{\sqrt{1+2n_r-2m_r\cos({\arg\lambda-\arg m_r})}}{\,\,\,1+2n_r+2m_r\cos({\arg\lambda-\arg m_r)}}~.
\end{dcases} 
\end{equation}
Note that when $\arg\lambda-\arg m_r\sim\pi/2$, it is $\xi\sim 0$ and the asymmetry between $p_a$ and $p_b$ is lost, even for large bath correlation strength $m_r\gg 1$.

\section{Fully connected models}
To bring out the correspondence between the spin systems and the considered nonlinear resonators, let us consider the Lipkin–Meshkov–Glick (LMG) model, which is a fully connected Ising model. For the sake of simplicity, we consider a single spin ensemble, dropping the superscripts $a$ and $b$,
\begin{equation}
\label{H_LMG}
\hat{H}_{\rm LMG} = h \sum_{i=1}^N \hat{\sigma}_i^z - \frac{\Lambda}{N} \sum_{i,j=1}^N \hat{\sigma}_i^x \hat{\sigma}_j^x =
h \hat{J}_z - \frac{\Lambda}{N} J_x^2.
\end{equation}
where we defined the collective spin operators $\hat{J}_z = \sum_i \hat{\sigma}_i^z$ and $\hat{J}_x = \sum_i \hat{\sigma}_i^x$. We divided the coupling term by $1/N$ so that the total energy remains extensive. As the LMG Hamiltonian commutes with the total angular momentum operator $\hat{J} = \hat{J}_x + \hat{J}_y + \hat{J}_z $, we can decompose the interaction term using raising and lowering operators in the highest angular momentum sector, that is $\hat{J}_x = (\hat{J}_+ + \hat{J}_-)/2$. Now, using the Holstein-Primakoff transformation, we can describe the spin excitations in terms of a collective bosonic mode such that $\hat{J}_z = \crea{a}\ann{a} - N/2$ and $\hat{J}_+^a =  \hat{a}^\dagger \sqrt{N-\hat{a}^\dagger \hat{a}}$, which can be expanded as 
$\hat{J}_+^a = \sqrt{N}\ \hat{a}^\dagger -\frac{1}{2 \sqrt{N} } \hat{a}^\dagger \hat{a}^\dagger \hat{a} +  \mathcal{O}(N^{-1})$. We can then write
\begin{align}
&\hat{J}_+^2 = N\ \crea{a}\crea{a} - \frac{1}{2}\round{ \crea{a}\crea{a}\crea{a} \ann{a} + \crea{a} \crea{a} \ann{a} \crea{a} } + \mathcal{O}(N^{-1})\\
&\hat{J}_+ \hat{J}_- = N \crea{a} \ann{a} - \crea{a}\crea{a} \ann{a} \ann{a} + \mathcal{O}(N^{-1}),
\end{align}
so that
\begin{align}
\frac{1}{N}\ \hat{J}_x^2 &= \frac{1}{4}\round{ \hat{J}_+^2 + \hat{J}_-^2 + \hat{J}_+\hat{J}_- + \hat{J}_-\hat{J}_+ }  =
  \frac{1}{4}\round{ \hat{J}_+^2 + \hat{J}_-^2 + 2\hat{J}_+\hat{J}_- - 2\hat{J}_z } + \mathcal{O}(N^{-2})  = \\
  &=\crea{a}\crea{a} + \ann{a}\ann{a} + 2\crea{a}\ann{a} - \frac{1}{N}\round{2\crea{a}\ann{a} + 2\crea{a}\crea{a}\ann{a}\ann{a}} 
 - \frac{1}{2N}\round{ \crea{a}\crea{a} + \ann{a}\ann{a} +2 \crea{a}\crea{a} \crea{a} \ann{a} + \crea{a}\ann{a}\ann{a}\ann{a}} + \mathcal{O}(N^{-2}) .
\end{align}
Up to terms of the order $\mathcal{O}(N^{-2})$, and up to constants, the full LMG Hamiltonian can then be rewritten as
\begin{equation}
\hat{H}_{LMG} = \round{h-\frac{\Lambda}{2} + \frac{\Lambda}{2N}}\crea{a}\ann{a} - 
\round{\frac{\Lambda}{4} - \frac{\Lambda}{8N}}\round{\crea{a}\crea{a} + \ann{a}\ann{a} } + 
\frac{\Lambda}{2N}\round{ 2\crea{a}\crea{a} \ann{a}\ann{a} + \crea{a}\crea{a}\crea{a}\ann{a} + \crea{a}\ann{a}\ann{a}\ann{a}  }.
\end{equation}
In the thermodynamic limit $N\to\infty$ and under the rotating-wave approximation, we can finally write
\begin{equation}
\hat{H}_{LMG} = \round{h-\frac{\Lambda}{2}} \crea{a}\ann{a} - \frac{\Lambda}{4} \round{\crea{a}\crea{a} + \ann{a}\ann{a} } 
+ \frac{\Lambda}{2N}\ \crea{a}\crea{a}\ann{a}\ann{a}.
\end{equation}
Accordingly, in the thermodynamic limit, up to orders of $1/N^2$, the Hamiltonian of Eq.~\eqref{H_LMG} can be mapped into the nonlinear resonators considered in section~\ref{sec_model_derivation}, by identifying $\omega = h-\frac{\Lambda}{2}$, $\lambda = \Lambda/2$ and $\epsilon = \Lambda/2N$. This derivation clarifies the one-to-one correspondence between the thermodynamic limit $N\to\infty$ and the scaling limit $\epsilon \to 0$.

\clearpage
\newpage
\bibliography{References}

\end{document}